\documentclass{aa}
\usepackage[varg]{txfonts}

\begin{document}
\title{Pan-Planets: Searching for hot Jupiters around cool dwarfs
\thanks{Based on observations obtained with the Hobby-Eberly Telescope, which is a joint project of the University of Texas at Austin, the Pennsylvania State University, Stanford University, Ludwig-Maximilians-Universität München, and Georg-August-Universität Göttingen.}
}

\author{C. Obermeier\inst{1,2,3}, J. Koppenhoefer \inst{2,3}, R. P. Saglia\inst{2,3}, Th. Henning \inst{1}, R. Bender\inst{2,3}, M. Kodric \inst{3,2}, N. Deacon \inst{4}, A. Riffeser\inst{3,2}, W. Burgett\inst{5}, K. C. Chambers\inst{6}, P. W. Draper\inst{7}, H. Flewelling\inst{6}, K. W. Hodapp\inst{6}, N. Kaiser\inst{6}, R.-P. Kudritzki\inst{6,3}, E. A. Magnier\inst{6}, N. Metcalfe\inst{6}, P. A. Price\inst{8}, W. Sweeney\inst{6}, R. J. Wainscoat\inst{6}, C. Waters\inst{6}}
\institute{{Max-Planck-Institute for Astronomy, Heidelberg, Königstuhl 17, D-69117 Heidelberg}
\and
{Max-Planck-Institute for Extraterrestrial Physics, Garching, Gießenbachstraße, D-85741 Garching}
\and
{University Observatory Munich (USM), Scheinerstraße 1, D-81679 Munich}
\and
{University of Hertfordshire, Hatfield}
\and
{GMTO Corp., 251 S. Lake Ave., Suite 300, Pasadena, CA 91101, USA}
\and
{Institute for Astronomy, University of Hawaii at Manoa, Honolulu, HI 96822, USA}
\and
{Department of Physics, Durham University, South Road, Durham DH1 3LE, UK}
\and
{Department of Astrophysical Sciences, Princeton University, Princeton, NJ 08544, USA}
}

\abstract{The Pan-Planets survey observed an area of 42 sq deg. in the galactic disk for about 165 hours. The main
scientific goal of the project is the detection of transiting planets around M dwarfs.
We establish an efficient procedure for determining the stellar
parameters $\mathrm{T_{eff}}$ and log\,$g$ of all sources using a method based on
SED fitting, utilizing a three-dimensional dust map and proper motion information. In this way we
identify more than 60\,000 M dwarfs, which is by far the largest sample
of low-mass stars observed in a transit survey to date.
We present several planet candidates around M dwarfs and
hotter stars that are currently being followed up. 
Using Monte-Carlo simulations we calculate the detection
efficiency of the Pan-Planets survey for different stellar and planetary
populations. We expect to find  $3.0^{+3.3}_{-1.6}$ hot Jupiters around F, G,
and K dwarfs with periods lower than 10 days based on the planet
occurrence rates derived in previous surveys. For M dwarfs, the percentage
of stars with a hot Jupiter is under debate. Theoretical models expect a lower occurrence rate than for larger main sequence stars. However, radial velocity surveys find upper limits of about 1\% due to their small sample, while the Kepler survey finds a occurrence rate that we estimate to be at least $0.17(^{+0.67}_{-0.04})$\%, making it even higher than the determined fraction from OGLE-III for F, G and K stellar types, $0.14(^{+0.15}_{-0.076})\%$. With the large sample
size of Pan-Planets, we are able to determine an occurrence rate of
 $0.11(^{+0.37}_{-0.02})$\% in case one of our candidates turns out to 
 be a real detection.
If, however, none of our candidates turn out to be true planets, we are able to put an upper limit of 0.34\% with a 95\%
confidence on the hot Jupiter occurrence
rate of M dwarfs. This limit is a significant improvement over
previous estimates where the lowest limit published so far is 1.1\%
found in the WFCAM Transit Survey.
Therefore we
cannot yet confirm the theoretical prediction of a lower occurrence rate for cool
stars. 
}
\date{Preprint online version: December 21, 2015}
\authorrunning{C. Obermeier et al.}
\titlerunning{Pan-Planets}
\maketitle
\section{INTRODUCTION}
As of July 2015, more than 1900 exoplanets have been discovered, the
majority of them with the transit method. One of the most noteworthy
discoveries, first detected with the 
radial velocity method, is the existence of hot Jupiters and hot Neptunes which
orbit closely around their host star. Such close-in gas giants
were unexpected since there is no equivalent in our solar
system. Those planetary systems are of significant interest, not
only for their unforeseen existence but also because they are the candidates best-suited for a planetary follow-up study with transit 
spectroscopy. Their large size lowers the difference between planetary and 
stellar radius. Besides the dependence on the atmospheric thickness, larger planetary radii improve the S/N of the transmission spectrum by increasing the overall surface area.  The radius ratio of hot Jupiters
and M-type dwarf stars is particularly favorable, although
only very few such systems have been detected so far \citep{2012AJ....143..111J,2014arXiv1408.1758H,2013A&A...551A..80T}. It is possible that they are  rarer than hot Jupiters around FGK stars since the amount of building material for planets is lower in M dwarf systems \citep{2005PThPS.158...68I,2010PASP..122..905J,2012A&A...541A..97M}. Additionally, there is a correlation between metallicity and giant planet occurrence rates for FGK stars \citep{1997MNRAS.285..403G,2001A&A...373.1019S,2005ApJ...622.1102F} with indications for the same correlation for M dwarfs \citep{2009ApJ...699..933J,2013A&A...551A..36N,2014ApJ...781...28M}. However, there is still an ongoing discussion about the strength of the metallicity dependence for M dwarfs \citep{2013ApJ...770...43M,2014ApJ...791...54G}.\\
Radial velocity (RV) surveys \citep{2007ApJ...670..833J,2013A&A...549A.109B} set an upper limit for the occurrence rate of hot Jupiters around M dwarfs of 1\%, however, with no precise estimates due to a small sample of a few hundred target stars per survey. These low sample sizes negate high detection efficiencies. 
\\Transit surveys, such as Kepler \citep{2012ApJ...753...90M,2013ApJ...767...95D,2014ApJ...791...54G,2014ApJ...791...10M,2015ApJ...807...45D}
and the WFCAM Transit Survey (WTS) \citep{2013MNRAS.433..889K,2013A&A...560A..92Z}, point to a fraction of less than 1\%
of M dwarfs that are being accompanied by a hot Jupiter. So far there have been few detections of such M-dwarf hot Jupiters \citep{2012AJ....143..111J,2014arXiv1408.1758H,2013A&A...551A..80T}. However, the sample sizes were not high enough to assess the occurrence rate accurately and all detected planets orbit only early M dwarfs.
\\Since radial velocity surveys provide information about the planetary mass and transit surveys about radii, it is not trivial to compare these results directly. Furthermore, many RV surveys focus on metal-rich host stars which seem to have a higher rate of hot Jupiters \citep{2013ApJ...767L..24D}. 
\\With Pan-Planets, we aim to address this issue by providing a substantially larger sample size. This survey has been made possible by the construction of a wide-field, high-resolution telescope, namely Pan-STARRS1 (PS1).
\\Pan-STARRS, the Panoramic Survey Telescope and Rapid Response System,
is a project with focus on surveying and identifying moving celestial bodies,
e.g. Near-Earth Objects that might collide with our planet.
The Pan-STARRS1 (PS1) telescope \citep{2002SPIE.4836..154K,2004AN....325..636H} is equipped with the 1.4 Gigapixel Camera
(GPC1), one of the largest cameras ever built. The size of the focal
plane is 40\,cm $\times$ 40\,cm which maps onto a 7 square field of
view \citep{2005AAS...20712101T,2009amos.confE..40T}. The
focal plane is constituted of 60 CCDs which are further segmented into
8\,$\times$\,8 sub-cells with an individual resolution of
\textasciitilde 600\,$\times$\,600 pixels at a scale of 0.258 arcsec per
pixel. A complete overview of the properties of the GPC1 camera can be
found in table \ref{tab:Pan-Planets}. The PS1 telescope is located at the
Haleakala Observatory on Maui, Hawaii. The central project of PS1 is an
all-sky survey that observes the whole accessible sky area of $3\pi$.\\ 
A science consortium of institutes in the USA, Germany, the UK and
Taiwan defined 12 key projects in order to make use of the large
amount of data being collected by the PS1 telescope. One of these key
projects is the dedicated Pan-Planets survey which has been granted
4\% of the total PS1 observing time. It began its science mission in May 2010.\\
With about 60000 M dwarfs in an effective FOV of 42 sq. deg., Pan-Planets is about ten times larger than previous surveys. In a sensitivity analysis of the project using Monte-Carlo simulations \citep{2009A&A...494..707K}, it was estimated that Pan-Planets would be able to detect up to dozens of Jovian planets that are transiting main sequence stars, depending on the observing time and noise characteristics of the telescope. The number of hot Jupiter detections around M dwarfs was undetermined since there was no reliable planetary occurrence rate. 
The actual photometric accuracy is lower than expected (see following Sect.) but good enough to detect transiting hot Jupiters around K and M dwarfs.\\
In Sect. \ref{survey} we describe the Pan-Planets survey and the data reduction
pipeline in detail. Our stellar classification and M dwarf selection is
presented in Sect. \ref{Mdwarf}. 
We detail our transit injection simulation pipeline that is being used for improved selection criteria and determination of the detection efficiency in Sect. \ref{simulation}.
An overview of the current survey status is given in Sect. \ref{status}. We detail the detection efficiency of the Pan-Planets
survey and discuss the results and implications in Sect. \ref{discussion}. Lastly, we draw our conclusions in Sect. \ref{conclusions}. \\
\section{SURVEY AND DATA REDUCTION}
\label{survey}
\subsection{Survey setup and execution}
\begin{figure}[htb]
\centering
\includegraphics[width=1\linewidth]{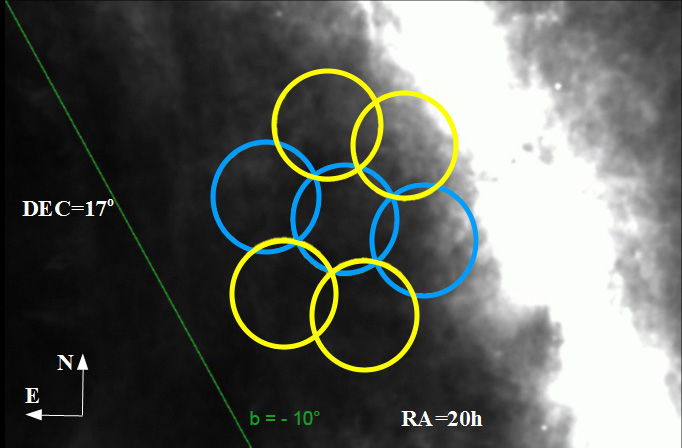}
\caption{Position of the Pan-Planets fields (coordinates in J2000). The yellow circles
  correspond to the four fields with data taken only in 2011 and
  2012. The blue circles correspond to the three fields with additional data
  taken in 2010. The background image is an
  extragalactic dust map taken from \citet{1998ApJ...500..525S}.}
\label{fig:dustmap}
\end{figure}

\begin{figure}[htb]
\centering
\includegraphics[width=0.8\linewidth]{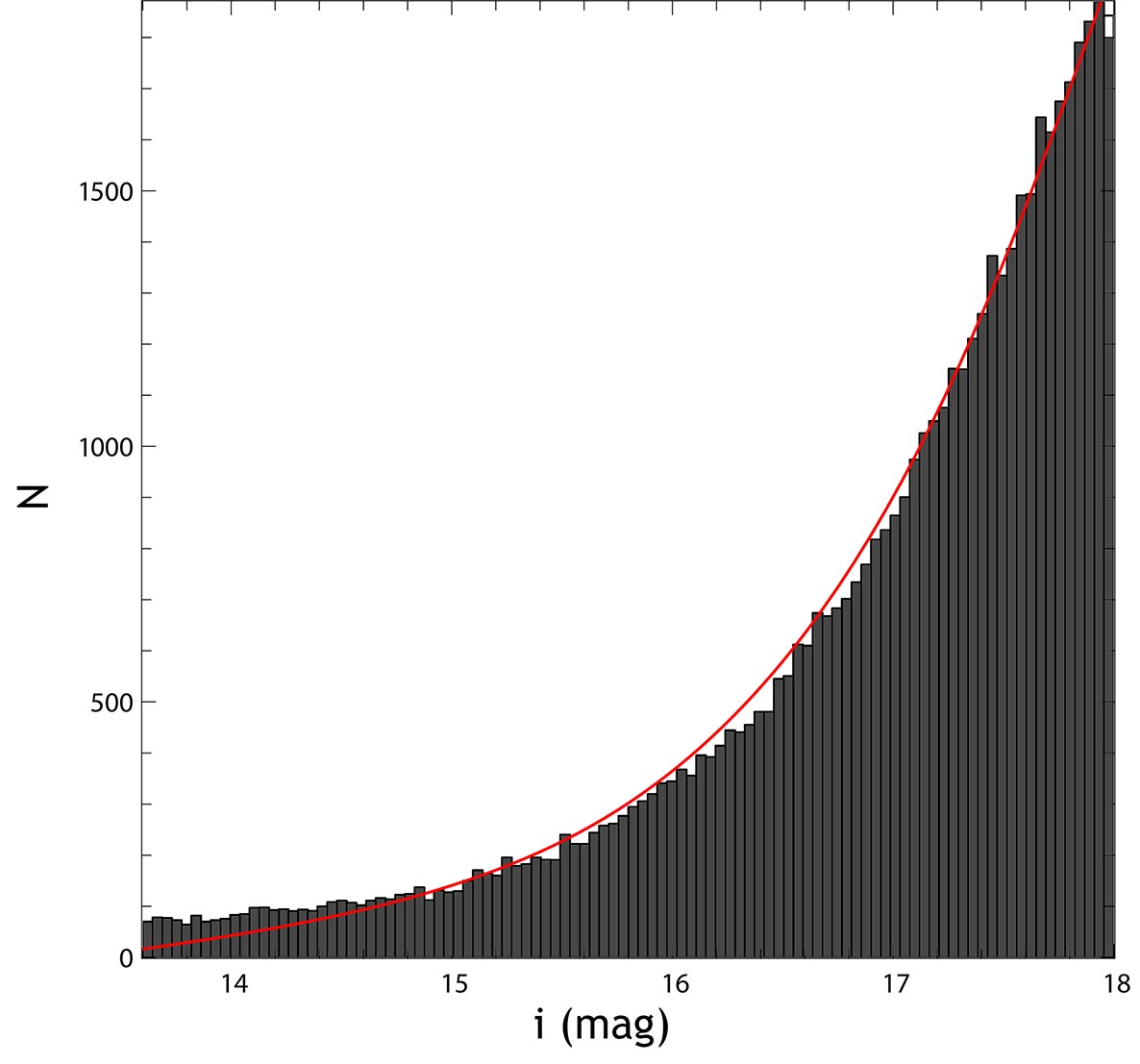}
\caption{Histogram with 100 bins of the brightness distribution in our 
M dwarf sample. The red line shows the distribution according to the Besan\c{c}on model \citep{2003A&A...409..523R}. 
Our fields include more bright stars than predicted by the Besan\c{c}on model, but the number of stars is in good agreement for stars with magnitudes $\mathrm{i' \geq 14.5\,mag}$.
We further detail our stellar classification method in Sect. \ref{Mdwarf}.}
\label{fig:Mdwarf-distribution}
\end{figure}

\begin{table}[!hbt]
\begin{tabular}{cc}
\hline\hline
\textbf{GPC specifications}\\\hline
Telescope          & 1.8m Pan-STARRS1 \\ 
Camera FOV         & 7 sq. deg.\\ 
Filters            & g', r', i', z', y'\\ 
Camera Properties  & 8x8 CCDs \footnotemark[1]\\ 
CCD Properties     & 8x8 cells with \textasciitilde 600x600 pixels\\ 
Pixel scale        & 0.258 arcsec/pixel\\ 
\\
\textbf{Pan-Planets characteristics}\\\hline
Observation period     & May 2010 - Sep. 2012 \\ 
Observation time       & 165 hours \\ 
Survey FOV             & 42 sq. deg.\\ 
Survey area            & 301.7$^{\circ}$ > RA > 293.7$^{\circ}$ \\ 
                       & 21$^{\circ}$ > DEC > 13$^{\circ}$ \\ 
Exposure time          & 15 or 30 s\\
& seeing-dependent\\
Median FWHM            & 1.07 arcsec \\ 
Photometric band       & i' \\
\#target stars           & \textasciitilde 4 $\mathrm{\cdot 10^6}$ \\
Target brightness      & 13.5 mag $\leq$ i' $\leq$ 18 mag \\
M dwarf targets        & \textasciitilde 60000\\ 
White dwarf targets    & \textasciitilde 4000\\ 
Observation time per night & 1 or 3 h \\ 
Photometric precision  & 5-15 mmag\\ \hline
\end{tabular} 
\caption{Properties of the GPC and the Pan-Planets survey.}
\label{tab:Pan-Planets} 
\end{table}
\footnotetext[1]{The GPC camera has a circular layout, i.e. the corners of the detector do not get illuminated.}
In 2010, Pan-Planets observed three slightly
overlapping fields in the direction of the Galactic plane. In the
years 2011 and 2012, four fields were added to increase the total
survey area to 42 square degrees in order to maximize the detection efficiency \citep{2009A&A...494..707K}. Fig. \ref{fig:dustmap} shows the
position of the seven Pan-Planets fields on the sky in relation to the extragalactic dustmap of \citet{1998ApJ...500..525S}.
\\Depending on atmospheric conditions, the exposure time was 30\,s or
15\,s. Observations were scheduled in 1h blocks. Over the three years of the project, we acquired 165\,h of observations, not including 15\,h from the commissioning phase in 2009.  The target
magnitude range of the survey is between 13.5 and 16.0 mag in the i'-band
which is expanded down to i'=18.0 mag for M dwarfs. The i' band is
ideally suited for a survey of cool stars, since those are relatively
bright in the infrared. Each field is split into 60 slightly
overlapping sub-fields (called skycells). The survey characteristics
of Pan-Planets are summarized in table \ref{tab:Pan-Planets}. More
information about the
planning of the survey can be found in \citet{2009A&A...494..707K}.\\
Focusing on stars smaller than the Sun has several advantages for the
search for transiting planets. The most significant one is that the
transit depth, which is the decrease in flux created by the planetary
transit, is determined by the square of the ratio between the planetary and stellar
radius. The smaller the star, the easier it is to detect the signal
since the light drop increases. This makes it possible to search for
hot Jupiters around very faint M dwarfs. Moreover, the M-dwarf stellar
type is the most abundant in our galaxy, meaning that there is a high
number of nearby cool dwarf stars, albeit very faint \citep{2006AJ....132.2360H,2015AJ....149....5W}. We estimate that
our sample contains up to 60000 M dwarfs (details on our stellar
classification can be found in Sect. \ref{Mdwarf}).  This M dwarf
sample is several times larger than in other transit surveys such as
Kepler or WTS, enabling us to determine the fraction of hot Jupiters
around M dwarfs more precisely. We show the brightness
distribution of our selected M dwarf targets in Fig. \ref{fig:Mdwarf-distribution}.\\
\subsection{Basic image processing}
\label{datareduction}
\begin{figure}[!ht]
\centering
\includegraphics[width=0.8\linewidth]{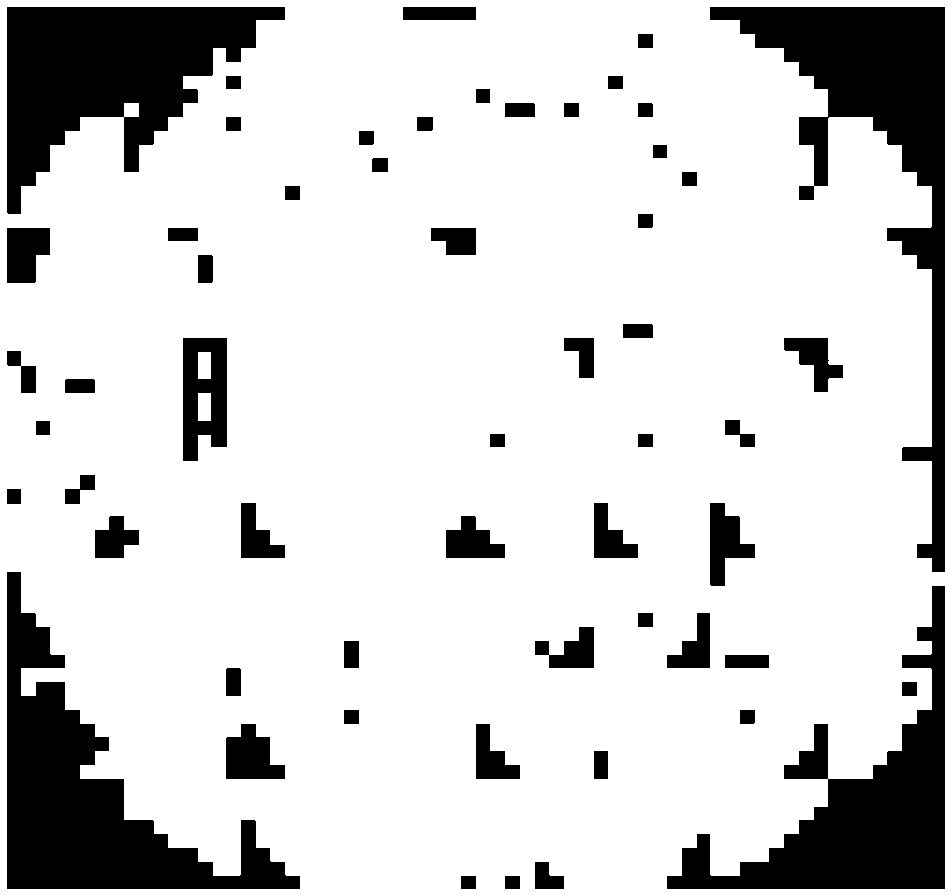}
\caption{Statically masked areas in the GPC1 camera for the 2012
  data. Note that the
    corners are not illuminated due to the circular layout of the GPC1 camera.}
\label{fig:Chipmask}
\end{figure}

\begin{figure}[!ht]
\centering
\includegraphics[width=1\linewidth]{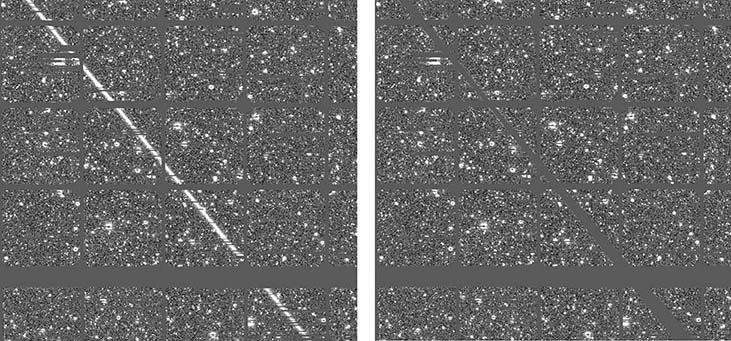}
\caption{Left: Satellite trail in one of the Pan-Planets
  images. Right: Result after automatic masking.}
\label{fig:Sat-comb}
\end{figure}

\begin{figure}[!ht]
\centering
\includegraphics[width=1\linewidth]{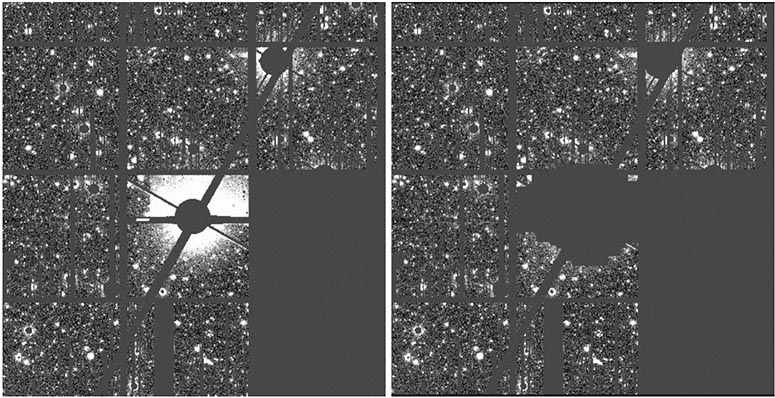}
\caption{Left: Saturated area that has not been sufficiently
  masked. Right: Result after application of the automatic
  masking. }
\label{fig:Bright}
\end{figure}

\begin{figure}[htb]
\centering
\includegraphics[width=0.9\linewidth]{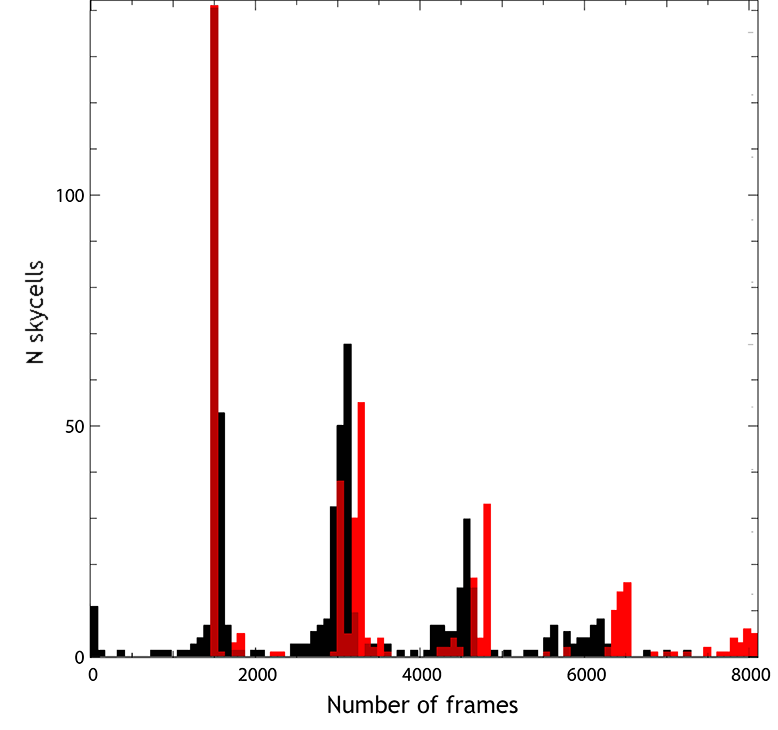}
\caption{Histogram of the number of frames per skycell, of which there are 420 overall. In red we
  show the distribution per skycell before ingesting the images into
  the pipeline, in black after ingesting. One can see that the smallest
  overlapping region completely vanishes and only a small fraction of
  skycells with about 6000 frames remains.}
\label{fig:hist-frames}
\end{figure}

All images have been processed in Hawaii by the PS1 Image Processing
Pipeline (IPP, \cite{2006amos.confE..50M}), which applies standard image
processing steps such as de-biasing, flat-fielding and astrometric
calibration. Each exposure is resampled into 60 slightly overlapping
sub-cells (skycells). Every skycell has a size of
$\sim$6000\,$\times$\,6000 pixels and covers an area of
30\,$\times$\,30 arcminutes on the sky.\\ During the analysis of the early data releases we realized that
several cells of the GPC1 CCDs (mostly located in the outer areas) exhibit a high level of systematics. To
account for that, we use time-dependent static masks that we created and provided to the IPP team (see Fig. \ref{fig:Chipmask} for the chip mask used for the 2012 data).
\\The re-sampled IPP output images have been transferred to Germany and
stored on disk for a further dedicated analysis within the
Astro-WISE\footnote{Astronomical Wide-field Imaging System for Europe, http://www.astro-wise.org/}
environment \citep{2013ExA....35....1B}. During the ingestion of the
data into Astro-WISE we correct for several systematic
effects. We apply an automated algorithm that searches
for and subsequently masks areas that display a systematic offset with respect to
the surrounding areas (e.g. unmasked ghosts, sky background
uniformities, etc.). Since satellite trails are not removed by the
IPP, we apply a masking procedure based on a Hough
transformation \citep{Duda:1972:UHT:361237.361242} that is available in
Astro-WISE. Fig. \ref{fig:Sat-comb} shows an example image before and
after the satellite trail masking.
\\Blooming of very bright stars is confined to one of the 8 $\times$ 8
cells of each chip. We apply an algorithm that detects saturated or overexposed areas and then masks the surrounding region,
as demonstrated in Fig. \ref{fig:Bright}.
\\Since the skycells are overlapping and three of the seven field have
been observed longer, the total number of frames per skycell is
varying between 1700 and 8400. Fig. \ref{fig:hist-frames} shows a
histogram of the number of frames per skycell.
\\Within one skycell there is significant masking in a majority of
the frames. Our data reduction pipeline discards any frame with less
than 2000 visible sources, which corresponds to masking of
about 85\%. The histogram of remaining frames is shown in
Fig. \ref{fig:hist-frames} in black. One can see that many images from
the overlapping regions with a high initial number of images (red)
are dropped due to high masking. The
comparatively low resulting number of frames, especially in the four
less visited fields, significantly influences the detection efficiency for 
planets with long periods or shallow transits.\\
There is a noticeable difference in data quality
between the 2010 data in comparison to the 2011 and 2012 data. In the
first year, the camera read-out resulted in a systematic astrometric shift of
bright sources (i' $\leq$ 15.5 mag) with respect to faint sources. This effect was noticed
in early 2011 and fixed by adjusting the camera voltages. In order to
account for the shifted bright stars, we use custom masks in our data
analysis pipeline for the 2010 data.\\
\subsection{Light curve creation}
\begin{figure}[htb]
\centering
\includegraphics[width=1\linewidth]{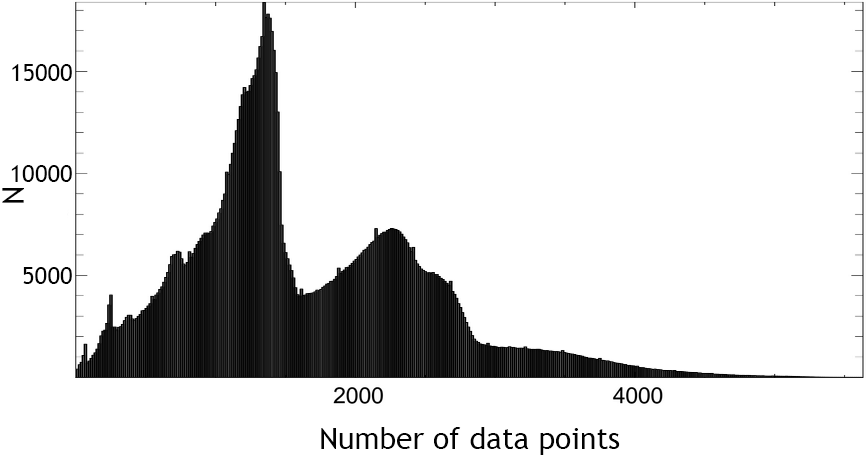}
\caption{Histogram of the number of data points per source. One can
  see two larger peaks, being created by the additional observations for 3 fields 
  in 2010. Overlapping regions, seeing-dependent exposure times and static
   masking of some detector areas broaden those peaks. The additional data from overlap further create a tail, reaching up to 5000 data points.}
\label{fig:datapoints}
\end{figure}

\begin{figure}[hbt]
\centering
\includegraphics[width=1\linewidth]{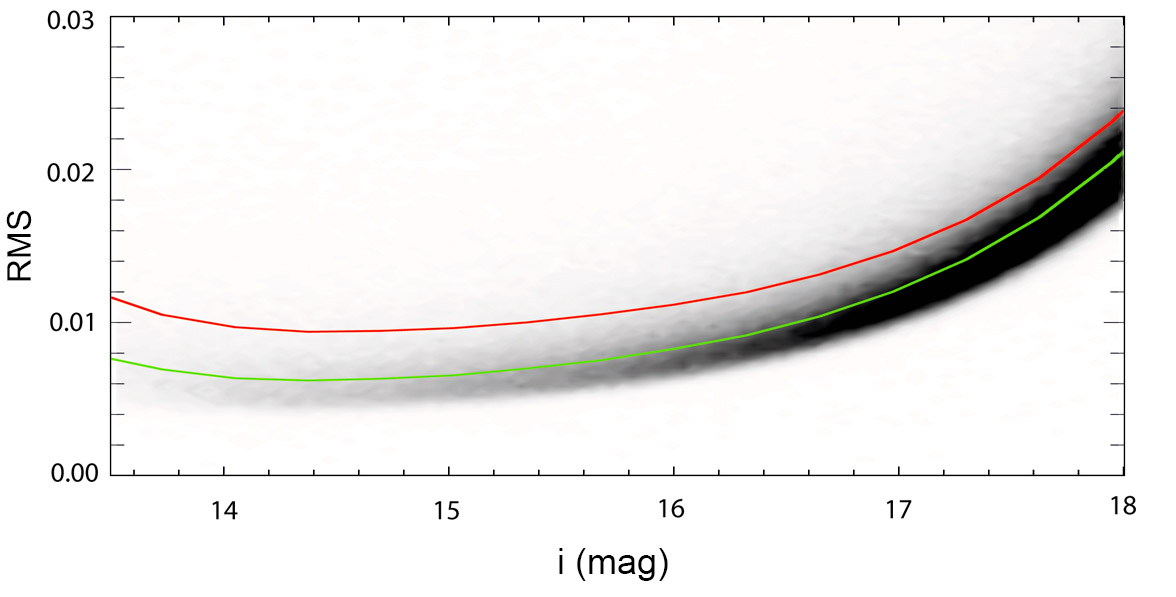}
\caption{Density plot of $RMS$ against the i-band magnitude in the central
  field after iterative clipping of 5$\sigma$ outliers and application
  of the \textsl{sysrem} algorithm. The green line shows the median values in
  0.2\,mag bins, the red line the values before application of
  \textsl{sysrem}. Note that the phase is shifted by 0.5 units for better readability. }
\label{fig:rms-mag-field5}
\end{figure}

The Pan-Planets light curves are created using the Munich Difference
Imaging Analysis (MDia) pipeline \citep{2013ExA....35..329K, 2002A&A...381.1095G}. This
Astro-WISE package makes use of the image subtraction method which
was developed by \citet{1996AJ....112.2872T} and later by
\citet{1998ApJ...503..325A}. The method relies on the creation of a
reference image, which is a combination of several images with
high image quality, i.e. very good seeing and low masking. As discussed in
\citet{2013ExA....35..329K}, increasing the number of input images
increases the S/N of the reference frame. However, each additional
image broadens the PSF which means that resolution decreases. Due to
the high masking in the Pan-Planets images (the average masking is
$\sim$40\% including cell gaps) we decide to use a high number of 100
input images, resulting in a typical median PSF FWHM of 0.7\,arcseconds in the 
reference frame.\\
The procedure to select the 100 best images is the following: after
removing all frames with a masking higher than 50\%, we select the 120
images with the best seeing.
We determine the weight of each image on the reference frame by determining the PSF FWHM and S/N and reject
frames that possess a very low weight (less than half of the median
weight) or too high weight (higher than twice the median weight), 
which usually results in 10 removed frames. This
is necessary in order to avoid using bad images that do not contribute in S/N or images that would
dominate the final reference frame, adding noise. Out of the remaining images we 
clip the frames with the broadest PSF until we have the final list for the best 100 frames.
These images are subject to a visual inspection in
which any leftover systematic effect is masked by hand before
combining the images to create the reference frame.\\
The next step is to generate the
light curves for each individual source. We photometrically align each
 image to the reference frame and correct for background and zeropoint
differences. Subsequently we convolve the reference image with a
normalized kernel to match the PSF of the single image and subtract
it. In the resulting difference image we perform PSF-photometry at
each source position. We calculate the total fluxes by adding the
flux measured in the difference images with the flux in the reference
image which is measured using an iterative PSF-fitting
procedure. Fig. \ref{fig:datapoints} shows a histogram of the number of datapoints for each source. One can see two broad peaks. The second peak, having more data points, is created by the additional observations for 3 fields that were taken in 2010. \\
Since the output light curves of MDia are at an arbitrary flux level,
we calibrate them by applying a constant zero point correction for each
skycell that is derived using the 3$\pi$ catalogue (version PV3) from Pan-STARRS1 as
a reference.
\\While analysing the light curves, we found that some data have a lower quality
depending on the time of observing. This applies mostly to the 2009 data for which a different camera configuration and survey strategy were
used. Hence, we decide to disregard the 2009 data for the further data reduction.
We perform an a-posteriori error bar correction on the light
curves. This is done by rescaling the error values of every light
curve using a magnitude-dependent scaling factor, derived by
fitting a forth-order polynomial to the magnitude dependent ratio
between the median error value and the RMS of each light curve.\\
To remove systematic effects that appear in many light curves, we apply
the \textsl{sysrem} algorithm that was developed by 
\citet{2005MNRAS.356.1466T}. The concept of \textsl{sysrem} is to analyse a 
large part of the data set, in our case one skycell, and identify
systematic effects that affect many stars at the same time. 
 The strength of this
algorithm is the fact that it does not need to know the cause of the
systematic effects it corrects. However, for \textsl{sysrem}
 to work properly, we have to remove stars with high intrinsic variability from
  the data sample beforehand. We do this by eliminating stars which have a
   reduced $\chi^2$ higher than 1.5 for a
constant baseline fit which subsequently also do not get corrected by $sysrem$. This way, we include about 80\% of the light
curves. Fig. \ref{fig:rms-mag-field5} shows the
overall quality of the light curves and the improvement that is
achieved by utilizing this algorithm, namely the $RMS$ scatter of the
Pan-Planets light curves as a function of i-band magnitude. At the
bright end we achieve a precision of $\sim$5\,mmag.\\
\subsection{Light curve analysis}
\label{signal detection}
\begin{figure}[hbt]
\centering
\includegraphics[width=1\linewidth]{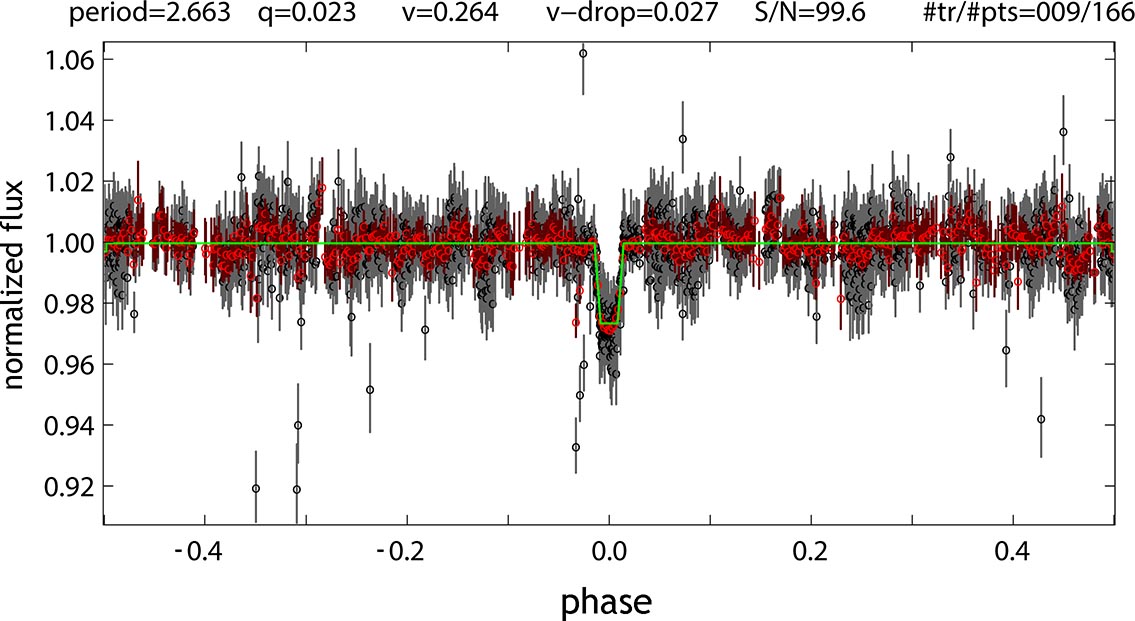}
\caption{Typical plot of our signal detection algorithm for object
 1.40\_14711, a K dwarf being orbited by a hot Jupiter candidate. Low
  resolution spectroscopy confirms the stellar type determined through SED
   fitting (see also Sect. \ref{Mdwarf}). Shown at the
  top are period (days), transit duration q (in units of phase),
  transit v shape (0 corresponds to a box, 1 to a V), transit light drop, S/N and number of
  transits/number of points in the transits. The binned data points
  are shown in red. A green line shows the best-fitting 2-level
  system, including the v-shape adjustment.}
\label{fig:transit-example}
\end{figure}

We search for periodic signals in the Pan-Planets light curves with an
algorithm that is based on the box-fitting least squares (BLS)
algorithm of \citet{2002A&A...391..369K}. It is very
efficient in detecting periodical signals which can be approximated by
a two level system, such as a planetary transit. We extend the BLS
algorithm by a trapezoid-shaped re-fitting at the detected periods, which we call transit v-shape fitting. A value of 0 corresponds to a box shape and 1 to a V.
It is a better representation of the true shape of eclipse
events. Further, we fit for a possible secondary transit, offset by 0.5
phase units, in order to discriminate between planets and eclipsing
binaries. Eccentric orbits are not uncommon for binaries, hence the secondary might also appear at different phases. Our BLS algorithm is not optimized to detect secondary eclipses at phases other than 0.5. 
Eclipsing binaries of interest (see Sect. \ref{cand}) that exhibit visible eccentricity will be analysed further with an adaptation of our BLS code. More detailed information on the modifications of our BLS
algorithm can be found in \citet{2013A&A...560A..92Z}.\\
We test 100001 periods distributed between 0.25 and 10\,days for each light
 curve. In order to speed up the fit we bin the phase folded light curve to 500 points, a number which we determined through dedicated Monte-Carlo simulations (same as in Sect. \ref{simulation}) in which we determined the effect of binning on the detection efficiency. The transit duration is limited to 0.25
phase units. This does not constrain the planet transit duration. As an extreme case. the duration for a hot Jupiter with a 12\,h period around an M5 dwarf would still be less than 0.05. For non-circular orbits, the duration can increase, however, hot Jupiters are generally on rather circular orbits. The highest transit duration in our candidate sample is about 0.07. A typical plot of our signal detection output is shown 
in Fig. \ref{fig:transit-example}.
\subsection{Transit recovery}
\label{transit-recovery}
Having completed the BLS run, we need to preselect the light curves with 
a possible signal before visual inspection due to our large sample.  We retain the four best BLS detection for each light curve, i.e. those having the highest S/N.
We remove results close to alias periods introduced by
the window function of the observing strategy by 
utilizing Monte-Carlo simulations (see Sect. \ref{simulation}).  Out of the remaining detections we select the best fit, i.e. the one with the lowest $\chi^2$ of the trapezoidal re-fit.  
\section{M DWARF SELECTION}
\label{Mdwarf}
The large amount of M dwarfs in our sample makes it unfeasible to perform a
spectroscopic characterization for every star. Instead, we utilize a
combination of photometric and proper motion selection criteria. 
Strong reddening in several of our fields is problematic when using colour cuts.
Distant giant stars can be misclassified as M dwarfs as well as hotter main
 sequence stars that appear cooler due to reddening. This kind of
  misclassification could lead to large uncertainties in our sample.
We therefore utilize the Spectral Energy Distribution (SED) fitting method which allows us to estimate the effective temperature of stars through fitting of synthetic SEDs to multi-band photometry and identify the best-fitting model for every star. We limit the issue of dust reddening by making use of a distance-dependent dustmap (see following Sect.).
\subsection{SED fitting}
A necessary assumption for SED fitting is that the model stars are physically accurate since any issue in the synthetic sample has a strong impact on the selection process. We use four synthetic stellar libraries for the fit. The first one is the Dartmouth isochrone model from \citet{2008ApJS..178...89D}.
This database provides values for stellar mass, luminosity, surface gravity, metallicity and effective temperature. We limit the grid to solar metallicity since we encountered similar issues as \citet{2013ApJ...767...95D}, getting an overabundance of high-metallicity results. Furthermore, we use the PARSEC stellar isochrones \citep{2012MNRAS.427..127B} which are based on the Padova and Trieste stellar evolution code. We choose the newest version that is improved for low-mass stars \citep{2014MNRAS.444.2525C}. In order to achieve improved results at the lower mass region, we include the most recent isochrones from \citet{2015A&A...577A..42B} and the BT-Dusty models \citep{2012RSPTA.370.2765A}.
 Our final sample contains 25880 model stars with ages of 1-13 Gyr, masses of 0.1-40.5 M$_\odot$, effective temperatures of 1570-23186 K and radii of 0.13-299.61 $R_\odot$.
\\The fit becomes more precise with a higher amount of photometric information. 
We use the Pan-STARRS1 $3\pi$ survey (version PV3) bands g', r', i', z'  and the
2MASS bands J, H and K and combine those catalogues by coordinate matching.
We decide not to include the PS1 y-band for our fitting process. 
Conversion into the PS1 photometric system is achieved with polynomial extrapolation of the z magnitude.
Therefore, adding the y band would provide no useful physical information for the fit but create a bias towards the z photometry. After merging we achieve completeness for 62\% of all stars. For the remaining 38\% we only have PS1 photometry. Most of the missing stars are saturated in 2MASS. 
For stars that are listed in 2MASS, we have full photometric information in all seven bands for the majority of them (94\%). We do not impose thresholds on the 2MASS quality flags.
In order to stay consistent with our stellar targets, we limit the brightness range to $\mathrm{13.5\,mag \leq i_{PS1} \leq 18\,mag}$ for this catalogue.
\\Our first step is to determine the best-fitting distance modulus for each isochrone and photometric band $x$. The $\chi^2$ value of the distance fit for a star with apparent brightness $m_x$, distance modulus $d$ and the absolute brightness $M_x$ of the synthetic isochrone star is described by following term:
\begin{equation}
\chi^2 = \sum\limits_{x} \frac{(M_x -m_x +d)^2}{e_x^2}.
\end{equation}
Here, $e_x$ is the error of the magnitude $m_x$. We assume no errors for $M_x$. In order to find the best-fitting distance, we need to determine the minimum of $\chi^2$, therefore we take the derivative, solve it for the distance and end up with:
\begin{equation}
d = \frac{\sum\limits_{x} \frac{m_x -M_x}{e_x^2}}{\sum\limits_{x} \frac{1}{e_x^{2}}}.
\end{equation}
In case of zero extinction, this would give us the best fit for the distance. However, dust reddening is a significant factor for a large part of our fields. In order to solve this problem, we make use of the 3D dust map provided by \citet{2015ApJ...810...25G}\footnote{available at http://argonaut.rc.fas.harvard.edu/}. It gives a statistical estimate for the amount of colour excess E(B-V) for any point in our field, in distance modulus bins of 0.5\,mag in the range between 4\,mag and 15\,mag. We therefore assign a reddening term $\mathrm{R(d) \cdot f_x}$ for every star with a given distance modulus d, reddening coefficient $\mathrm{f_x}$ and photometric band x. We determine the reddening coefficients for each band through the web service NASA/IPAC Extragalactic Database (NED)\footnote{https://ned.ipac.caltech.edu/}, substituting the UKIRT J, H and K values for the 2MASS filters, using the dust estimates from \citet{2011ApJ...737..103S}. 
\\Fitting with a step function-like dust distribution results in artefacts. A first-order linear interpolation leads to similar, albeit weaker, artefacts. We therefore smooth by fitting a 10-th order polynomial to the points. This way, the distribution is artefact-free. 
With the given reddening R(d) for the best-fitting distance modulus d, we iterate the fit until the converging criterion
\begin{equation}
\mathrm{d_{difference} = |d_{n+1}-d_{n}| \leq 10^{-4}}
\end{equation}
is fulfilled.
This procedure is executed for each isochrone, after which we select the
best fit based on the lowest $\chi^2$ value. We interpolate missing error values in 2MASS by first fitting a magnitude-dependent polynomial to each band and then assigning the value for the given magnitude.
When comparing the $\chi^2$ values in relation to the measured distances, we 
find that there are usually two distinct local minima. This is explained by 
fitting two different stellar populations, e.g. main-sequence and giant 
branch. The resulting local minima sometimes show a very similar $\chi^2$, 
which makes it difficult - in those cases - to distinguish between different 
stellar populations. In order to solve this, we include proper motion 
information (see following subsection) into the classification.
\subsection{Proper motion selection}
Proper motion, in short PM, quantifies the angular movement of a star in the 
course of time from the observer's point of view. This is strongly 
correlated with the distance: the closer a star is to the observer, the 
higher (on average) the angular motion. Therefore we can be confident that 
if a star exhibits a high proper motion, the fit for the close distance is 
the most plausible one.
\\For this we utilize a combination of the USNO-B digitization of 
photometric plates \citep{2003AJ....125..984M}, 2MASS \citep{2006AJ....131.1163S}, the WISE All-Sky Survey 
\citep{2010AJ....140.1868W} and the 3$\pi$ Pan-STARRS1 
survey\footnote{http://ipp.ifa.hawaii.edu/} as described in \citet{2015arXiv150904712D}. 
After calculating the annual proper motion, we assign each star a quality 
flag depending on the properties shown in table \ref{PM}.
We select every cool star with quality flag 1, even if the best fit is 
slightly in favour of a distant red giant, and cool stars with the best fit 
for a dwarf type with quality flags 2, 4 and 5. 
\begin{table}[hbt]
\begin{tabular}{c|cc}
\hline\hline
Quality flag & PM value & Error  \\\hline
1 &  PM $\geq 6$ $mas/yr$ &  $PM_{error}/PM \geq 0.5$ \\ 
2 & PM $\geq 6$ $mas/yr$ &$PM_{error}/PM < 0.5$\\ 
3 & PM $< 6$ $mas/yr$ &$PM_{error}/PM \geq 0.5$\\ 
4 & PM $< 6$ $mas/yr$ &$PM_{error}/PM < 0.5$\\ 
5 &  No coordinate match. &\\ \hline
\end{tabular} 
\caption{Quality flags for different proper motion PM (mas/yr). Stars having flag 3 do not pass our criteria, having no measurable proper motion.}
\label{PM}
\end{table}
\\We further use the criterion $\mathrm{J-K > 1}$ as a flag to discriminate likely 
background giants from closer dwarf stars. This has proven to be very 
effective in the Kepler project \citep{2012ApJ...753...90M}. 
\subsection{Consistency check with the Besan\c{c}on model}
\label{results}
We compare our results to the Besan\c{c}on model \citep{2003A&A...409..523R} 
which provides a synthetic stellar
population catalogueue for any given point of the sky. We simulate our entire FOV in 1 sq. deg. bins. We use this to
estimate the distribution of spectral types in our target brightness. Choosing 
the criteria of an effective temperature < 3900\,K and surface gravity > 4, 
we identify 62800 M dwarfs in the Besan\c{c}on model.
We select M dwarfs in our survey with the following 
criteria:
\begin{itemize}
\item SED fitting Temperature < 3900\,K
\item Quality flag of 1 OR \\ 2, 4, 5 and a best fit for a nearby dwarf star
\end{itemize}
With those criteria, we select 65258 M dwarfs in our FOV (about 12000 M dwarf per field since there are multiple identifications in the overlapping regions). 
This is fairly consistent to the number of M dwarfs in the Besan\c{c}on model, however, our result is slightly higher. It is possible that there are false positive identifications in the selection list, reddened by dust from the galactic disc. This most likely affects identifications without proper motion data, i.e. flags 2, 4 and 5. However, the difference between our selection and the model distribution is not very large, so the amount of contamination is low.
\begin{figure}[!ht]
\centering
\includegraphics[width=\linewidth]{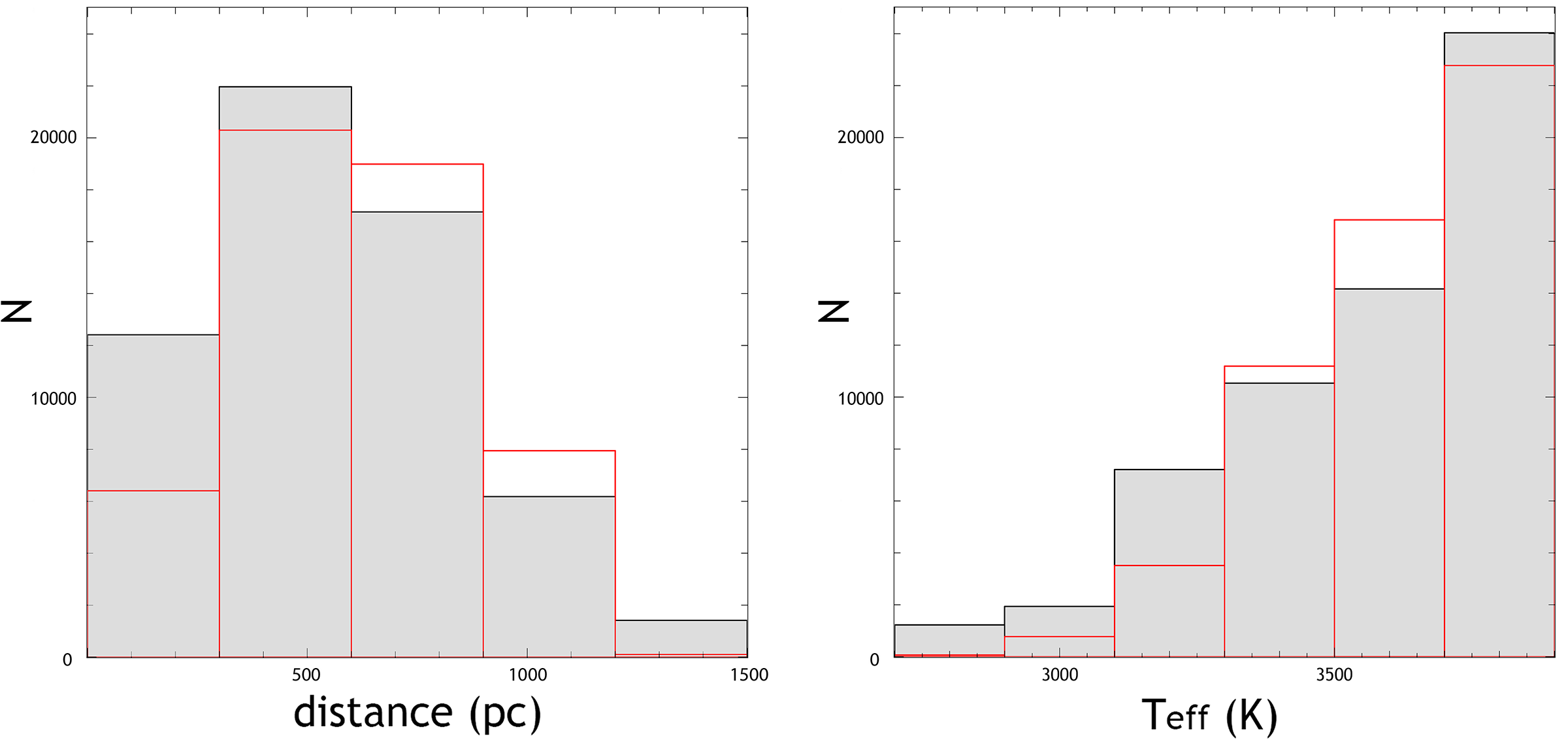}
\caption[Besan\c{c}on-T]{Left: distribution of distances for selected M dwarfs from SED fitting (gray with black bar lines) and the Besan\c{c}on model (red bar lines). Right: distribution of effective temperatures for selected M dwarfs from SED fitting (gray with black bar lines) with the expected distribution from the Besan\c{c}on model (red bar lines).}
\label{fig:Besandon-comp-T}
\end{figure}
\\Fig. \ref{fig:Besandon-comp-T} displays our implemented M dwarf selection and how it compares to the Besan\c{c}on model. The effective temperatures are fairly consistent, assuming an uncertainty of $\pm$100\,K for SED fitting. The distribution of distances does not seem to fit so well as the temperatures, however, we are mainly focused on fitting the effective temperature.  
\begin{figure}[!ht]
\centering
\includegraphics[width=\linewidth]{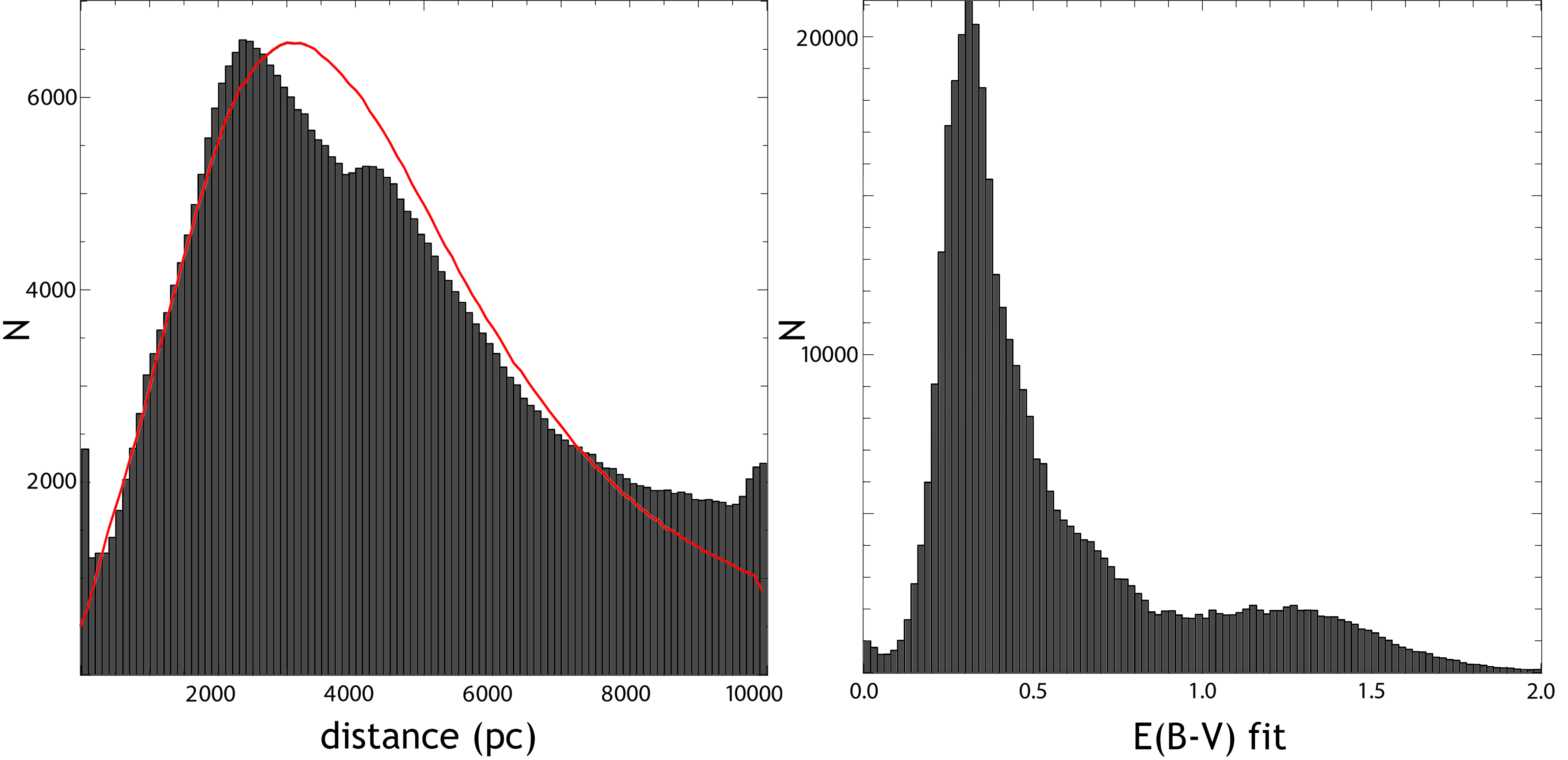}
\caption[hist-dist-ext]{Left: distribution of distances for all fitted stellar types, with the Besan\c{c}on model as a comparison (red line). Right: distribution of fitted extinction in our field.}
\label{fig:hist-dist-ext}
\end{figure}
\\Fig. \ref{fig:hist-dist-ext} shows the distribution of distances for all fitted stars in comparison to the Besan\c{c}on model. It seems to be very consistent in closer ranges, both distributions having their peak around 3\,kpc, but there are small divergences in the occurrence of distant ($\mathrm{\geq 3\,kpc}$) stars.  We are focused on nearby main-sequence stars so this is not much of a concern. In the same Fig. one can see the distribution of fitted extinction E(B-V) which peaks around an E(B-V) value of 0.4.
\begin{figure}[!ht]
\centering
\includegraphics[width=\linewidth]{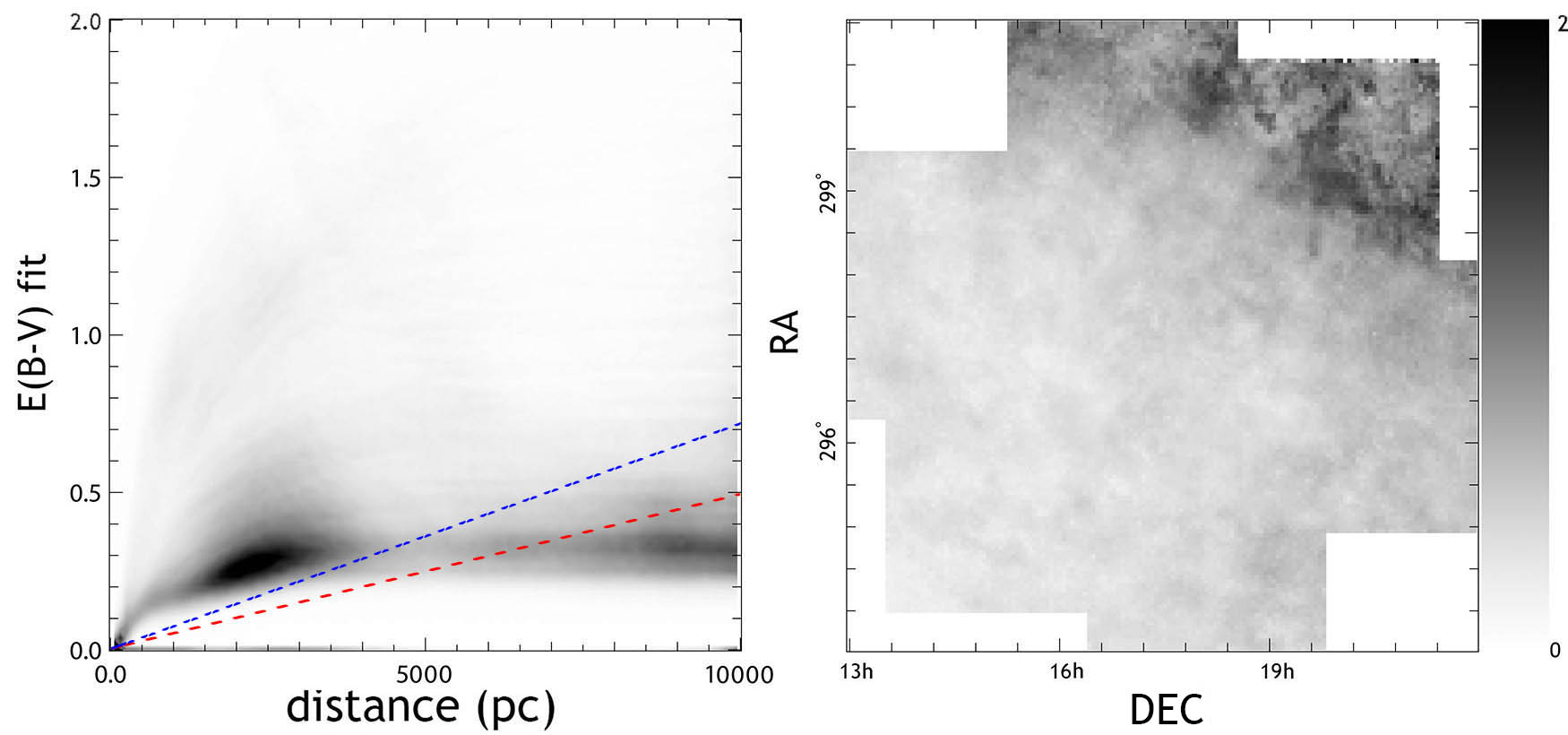}
\caption[plot-dust-dist]{Left: fitted reddening E(B-V) against distance for all fitted stars. As a comparison,  linear extinctions of 0.7\,mag/kpc  (red) and 1\,mag/kpc in the V-band as used in the Besan\c{c}on model and in \citet{2013ApJ...767...95D}, respectively, are overplotted as dashed lines. Right: average fitted E(B-V) in relation to the coordinates (J2000). One can see the dust-rich region in the upper right which is closer to the galactic disc.}
\label{fig:plot-dust-dist}
\end{figure}
\\In Fig. \ref{fig:plot-dust-dist} on the left is shown the relation between distance and reddening E(B-V) for all fitted stars while the right side of Fig. \ref{fig:plot-dust-dist} shows the average reddening in relation to the coordinates. As a comparison, we overplot the linear extinction models of 0.7\,mag/kpc as used in the Besan\c{c}on model (red) and 1\,mag/kpc as used in \citep{2013ApJ...767...95D} (blue). There is a noticeable difference for distances below 3\,kpc to our fitting. The outliers with an E(B-V) of more than 1.0 are due to the dust-rich region close to the galactic disc, shown in Fig. \ref{fig:plot-dust-dist} on the right (top-right corner).   
\subsection{Consistency check with Kepler targets}
As another consistency check, we take the SED fitting results used for 31 Kepler candidate M-dwarf host stars \citep{2013ApJ...767...95D}, identify the stars in the Pan-STARRS 3$\pi$ catalogue and perform SED fitting. In order to make the process more comparable, we limit our isochrones to less than solar masses, temperatures lower than 7000\,K and run the comparison with their model of extinction fitting, i.e. 1 mag in the V-band per kpc. 
\begin{figure}[!ht]
\centering
\includegraphics[width=0.9\linewidth]{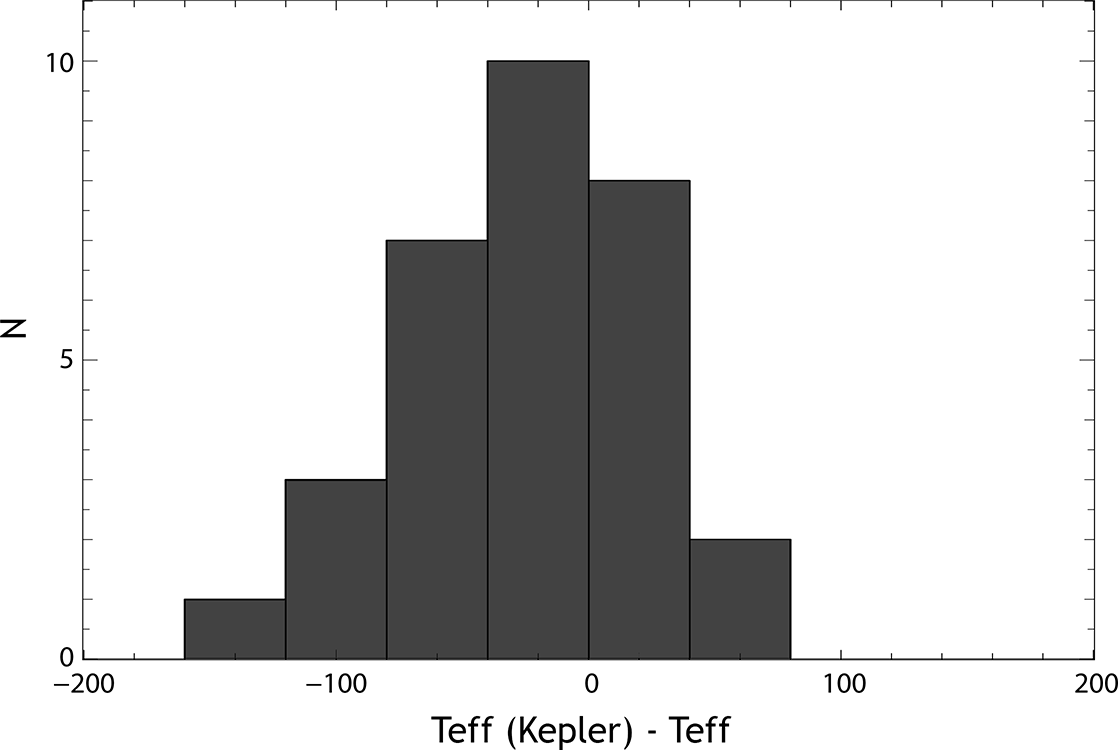}
\caption[Kepler-diff]{Difference of the calculated effective temperature between our SED fitting results and those of \citet{2013ApJ...767...95D}.}
\label{fig:Kepler-diff}
\end{figure}
\\As one can see in Fig. \ref{fig:Kepler-diff}, the results are fairly consistent but at the same time there is a systematic offset of about -25\,K. A likely explanation is that the fitting results from \citet{2013ApJ...767...95D} are for slightly older and therefore cooler stars. Additionally, the inclusion of non-solar metallicities might also explain or contribute to the shift.
However, the difference is not very large. We estimate to have an uncertainty of about $\pm100\,K$ which is larger than the observed difference. 
\subsection{Consistency check with spectroscopically confirmed M dwarfs}
\begin{figure}[!ht]
\centering
\includegraphics[width=0.9\linewidth]{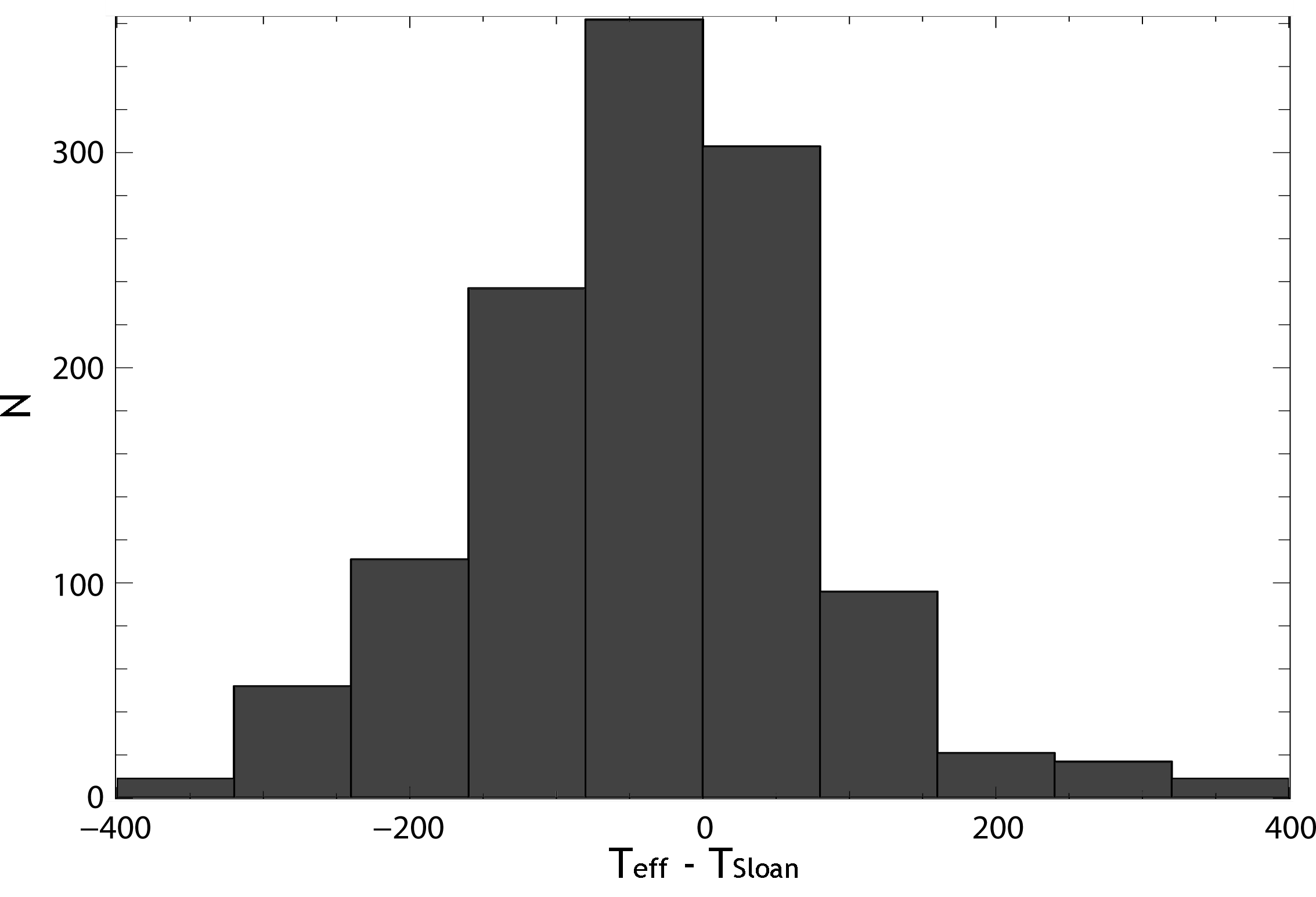}
\caption{Difference in fitted temperature between SED fitting and spectral fitting of the Sloan catalogue. Note that the temperature estimates from Sloan are in 200\,K bins, hence a scatter of $\mathrm{\pm 200\,K}$ is to be expected.}
\label{fig:Sloan-comp}
\end{figure}

As the final consistency check, we arbitrarily select 1000 confirmed M dwarfs out of the Sloan Digital Sky Survey Data Release 7 Spectroscopic M Dwarf Catalog (SDSS DR7) \citep{2011AJ....141...97W} that  
\begin{itemize}
\item exist in the PS1 3$\pi$ catalogue
\item exist in the 2MASS catalogue
\item have distance-dependent extinction data from \citet{2015ApJ...810...25G}
\item have data in all 7 bands
\item fit our target brightness range (13.5 $\leq$ i' $\leq$ 18).
\end{itemize}
This way we can make sure that the comparison is as close as possible as we use our regular stellar characterization pipeline.
We find that all of the listed M dwarf candidates are being identified as M dwarfs. Unfortunately, the effective temperatures of the SDSS DR7 catalogue are given in 200\,K bins, meaning that there is an inherent error of $\mathrm{\pm 100\,K}$ when comparing their estimates to ours. However, as is shown in Fig. \ref{fig:Sloan-comp}, there is very good agreement in the characterized temperatures between both methods. 
\subsection{SED fitting example} 
\label{SED-example}
\begin{figure}[!ht]
\centering
\includegraphics[width=\linewidth]{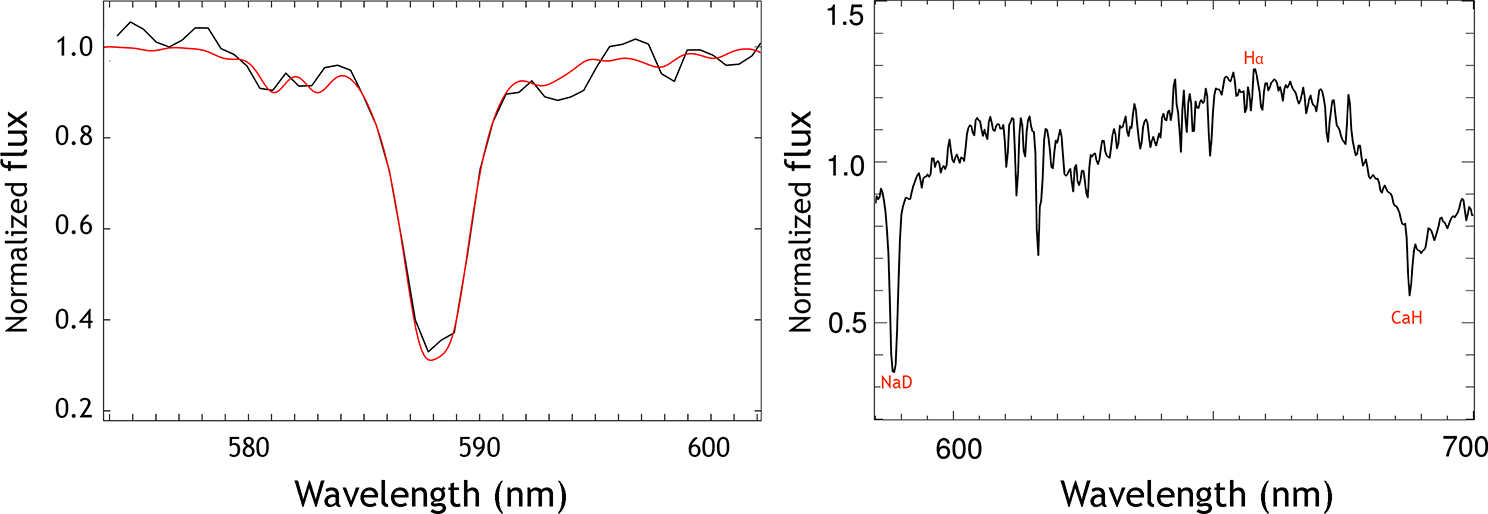}
\caption{Left: Best spectral fit (red) of the Na I line taken with the Otto Struve 2.1m ES2 low resolution spectrograph (black). Right: Complete spectrum recorded with the ES2 spectrograph with spectral features marked in red.}
\label{fig:Mdwarf-cand1-SED}
\end{figure}

\begin{figure}[!ht]
\centering
\includegraphics[width=\linewidth]{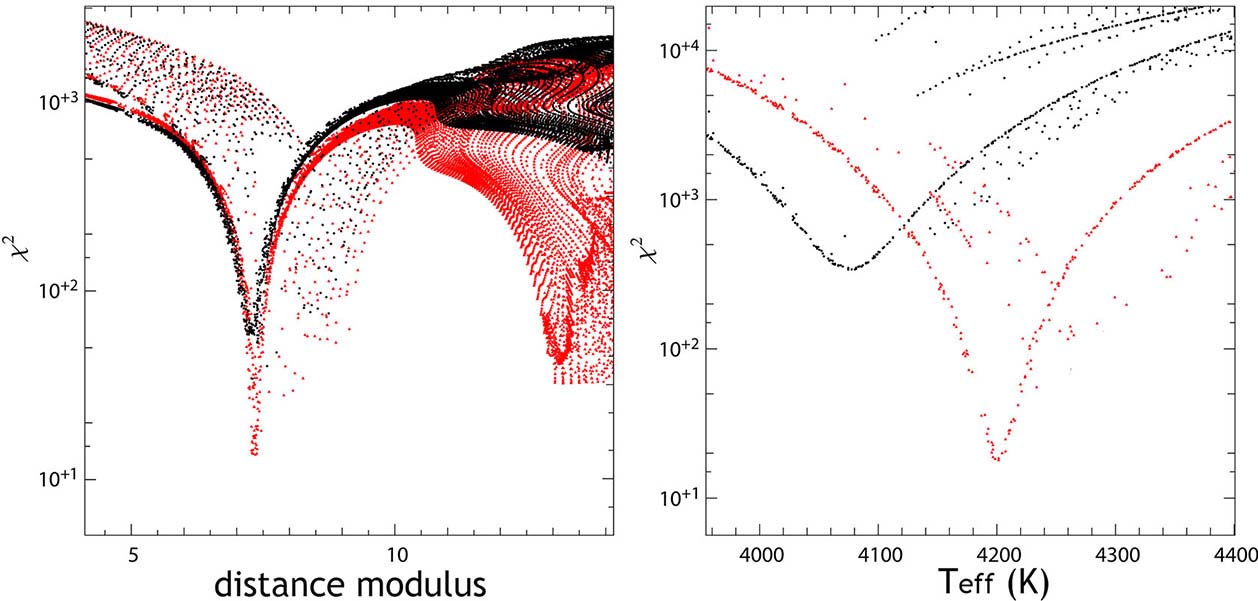}
\caption[cand1-SED]{Left: $\chi^2$ vs. distance modulus for hot Jupiter candidate system 1.40\_14711 with our implemented version of dust fitting (red), compared with a fit without extinction fitting (black). Right: $\chi^2$ vs. effective temperature for the same system with (red) and  without (black) extinction fitting. It is clear that SED fitting considerably improves $\chi^2$ and that there is no other alternative fit with an equally good fit.}
\label{fig:Mdwarf-cand-SED}
\end{figure}

To further confirm our SED fitting method, we recorded low resolution spectra for all 18 planet candidate stars with the Otto Struve 2.1m Electronic Spectrograph 2 (ES2) - a low resolution spectrograph. 
We illustrate the results for SED fitting and spectroscopy for candidate 1.40\_14711, a clear example of a late K dwarf. The candidate's light curve is shown in Fig. \ref{fig:transit-example}, in Fig. \ref{fig:Mdwarf-cand1-SED} its spectrum. 
\\With the spectrum we can confirm that the primary in this system is in fact an K dwarf at the boundary to the M dwarf regime, with strong NaD absorption around 5900~\AA, broad CaH absorption bands and very weak absorption in the H$\alpha$ band at 6563~\AA. The best fit of the surface gravity sensitive Na I doublet \citep{2012ApJ...753...90M} is shown in Fig. \ref{fig:Mdwarf-cand1-SED} on the left. The best result from spectroscopy is a star with log(g) of 4.5 and an effective temperature of between 4000\,K and 4250\,K, which is in good agreement with SED fitting. We further used the gravity sensitive region of 6470-6530\,\AA\ which are dominated by Ba II, Fe I, Mn I and Ti I lines \citep{1993PASP..105..693T}.
The SED fitting results for this candidate are shown in Fig. \ref{fig:Mdwarf-cand-SED}, with a best-fitting distance of 293\,pc and effective temperature of 4208\,K.  It is quite obvious that there is no alternative result that would have an equally good fit, e.g. a distant K or G giant reddened by dust. One can also see that our extinction fitting shifts the best-fitting temperature by about 125\,K. 

 \section{TRANSIT INJECTION SIMULATIONS}
 \label{simulation}
 The primary purpose of Pan-Planets is the detection of transiting hot Jupiters while setting new boundaries for the occurrence rate of close-up Jovian planets around M dwarfs. In order to do that, we need to determine the detection efficiency of this project. We perform extensive Monte-Carlo simulations, injecting planetary transit signals into the Pan-Planets light curves and trying to recover the signal. This is similar to other recent approaches performed on Kepler data, e.g. in \citet{2013PNAS..11019273P,2013ApJ...770...69P}, \citet{2015ApJ...810...95C} or \citet{2015ApJ...807...45D}. However, we utilize our full signal detection pipeline instead of inferring successful detections from calculating the number of visible transits combined with noise and signal to noise estimates. Our approach is much more suited to the peculiarities of Pan-Planets. A varying amount of data points, strong constraints for observational window functions and not well-defined systematics mean that this is the only reliable way of estimating our detection efficiency. 
 \subsection{Setup}
\begin{figure}[hbt]
 \centering
 \includegraphics[width=1\linewidth]{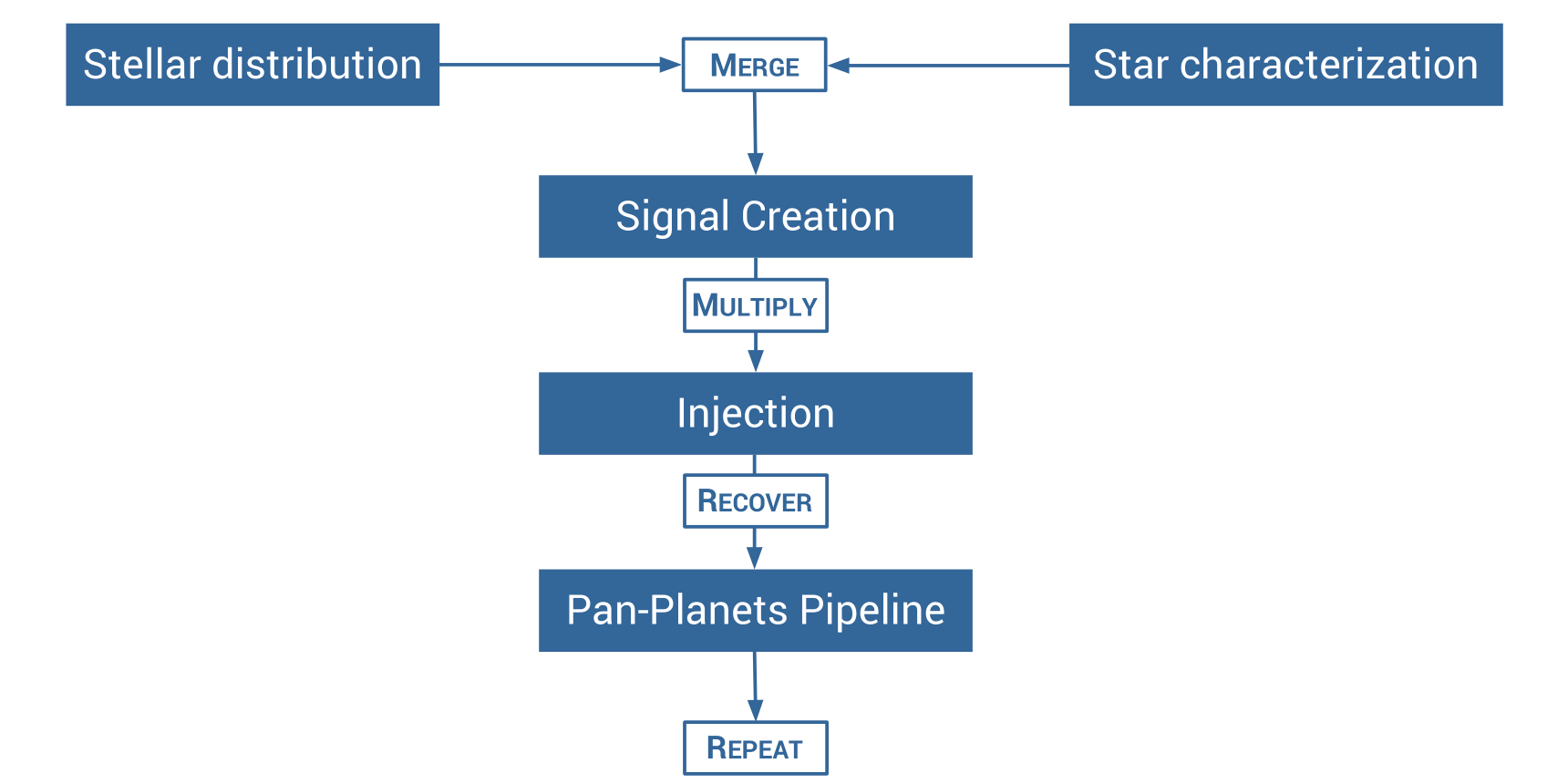}
 \caption{Illustration of our simulation process. We take the model stellar distribution and assign each of our characterized star to the closest-fitting model. We create a planetary signal out of the given stellar and planetary parameters, multiply it with the light curve and try to recover the injected transit with our Pan-Planets pipeline.}
 \label{fig:sim}
 \end{figure}

 We start by selecting all previously identified M dwarf light curves minus the identified planetary candidates. 
 We create a simulated distribution of different stellar parameters based on the Besan\c{c}on model \citep{2003A&A...409..523R} for our FOV. Each model star is assigned a set of real light curves, based on brightness. 
 The whole process is illustrated in Fig. \ref{fig:sim}. 
 \\In the next step, we create our target planet population by setting up random distributions of period and radius in defined boundaries. In accordance with \citet{2009ApJ...695..336H}, \citet{2009A&A...494..707K} and \citet{2013A&A...560A..92Z}, we use five different populations: Jovian planets with radii 1.0-1.2 R$\mathrm{_J}$ and periods of 1-3 days, 3-5 days and 5-10 days plus Saturn-sized and Neptune-sized populations with periods of 1-3 days and radii of 0.6-0.8 R$\mathrm{_J}$ and 0.3-0.4 R$\mathrm{_J}$, respectively. 
 \\We then take every star in the stellar distribution, randomly pick one of the corresponding light curves and select an arbitrary planet out of the chosen population. We assign a random geometrical inclination of the planetary orbit to each star and calculate the criterion 
 \begin{equation}
 \label{eq:sim}
 sin(i)< \frac{R_{star}+R_{planet}}{a},
 \end{equation}
 where $i$ is the inclination and $a$ the distance to the star. For simplicity, we assume a circular orbit. If this criterion is met, we create a transit signal based on all given parameters, i.e. planetary and stellar radii, inclination, period, t$_0$ and corresponding limb darkening coefficients \citep{2011A&A...529A..75C} for the stellar type. We multiply the simulated signal with the real light data and end up with simulated light curve that possesses all the characteristics of our survey, e.g. noise, systematics, distribution and amount of data points. Further information about the transit injection method used can be found in \citet{2009A&A...494..707K}. 
 \subsection{Transit recovery}
 \label{sec:trans-rec}
\begin{figure}[h]
\centering
\includegraphics[width=\linewidth]{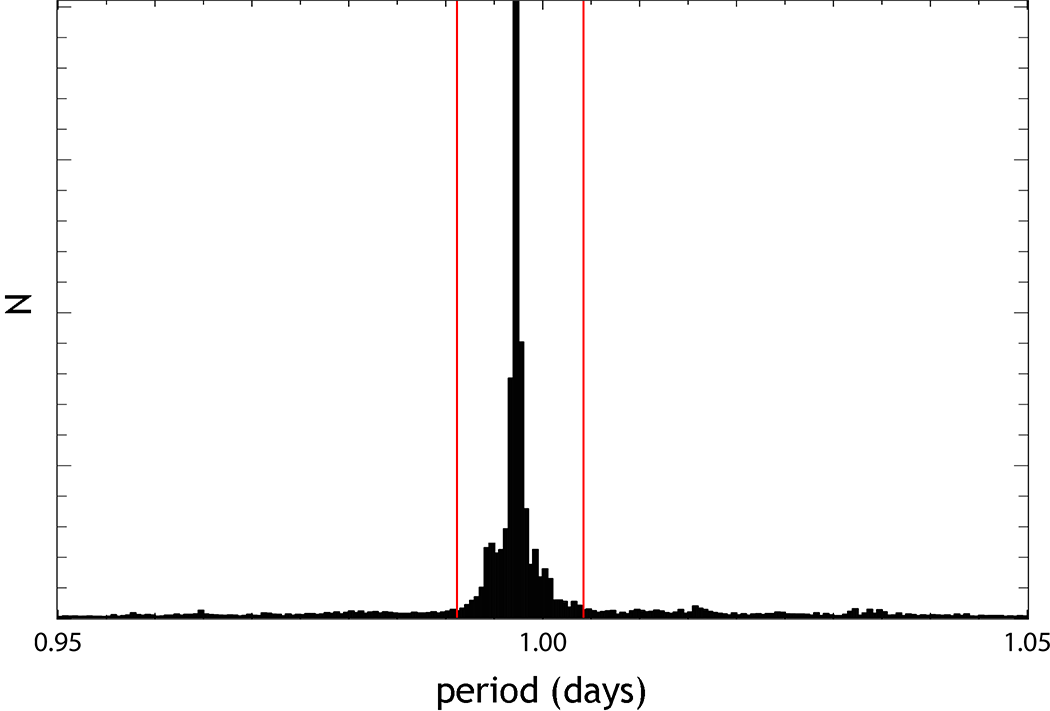}
\caption[alias-period]{Alias period around 1 day for an arbitrary number of hot Jupiter simulation runs, comprising the whole simulated period range of 1 to 10 days. The red lines mark the excluded period range from table \ref{tab:alias}. Due to window functions of the survey, the peak is
not directly at period 1.0 days, but slightly shifted to the left.}
\label{fig:alias-period}
\end{figure}

\begin{figure}[h]
  \centering
  \includegraphics[width=0.9\linewidth]{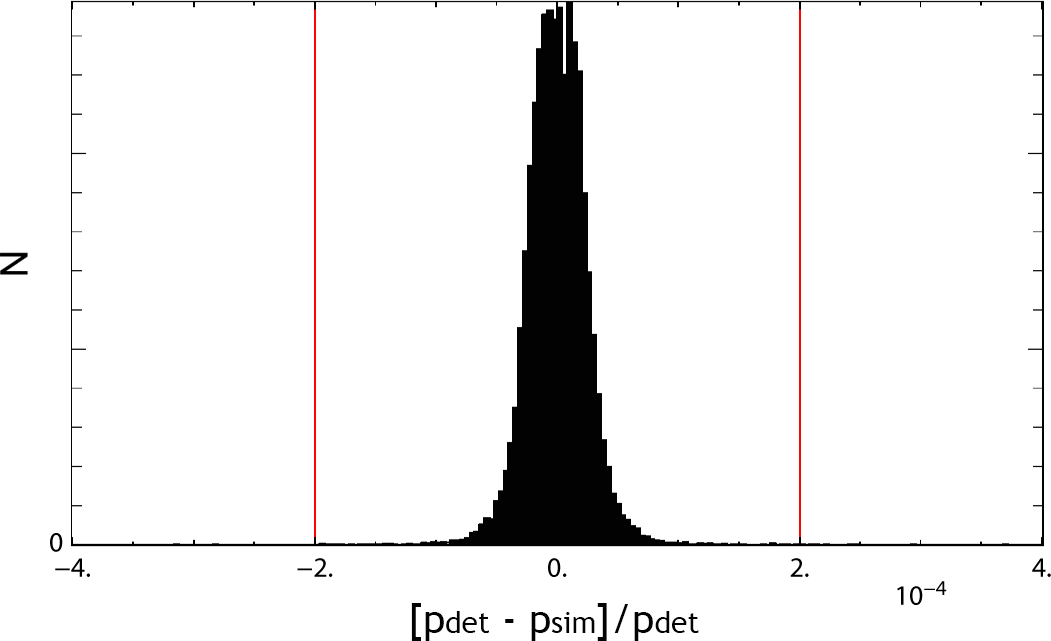}
  \caption[p-sim]{Deviation of the detected period p$_{det}$ from the simulated period p$_{sim}$ for an arbitrary number of hot Jupiter simulation runs, comprising of the whole field of view with periods between 1 and 3 days.
  For a successful detection, we require the detected period to deviate by a factor of less than 0.0002 from the simulated one (red lines).}
  \label{fig:hist-p}
  \end{figure}

\begin{figure}[h]
  \centering
  \includegraphics[width=\linewidth]{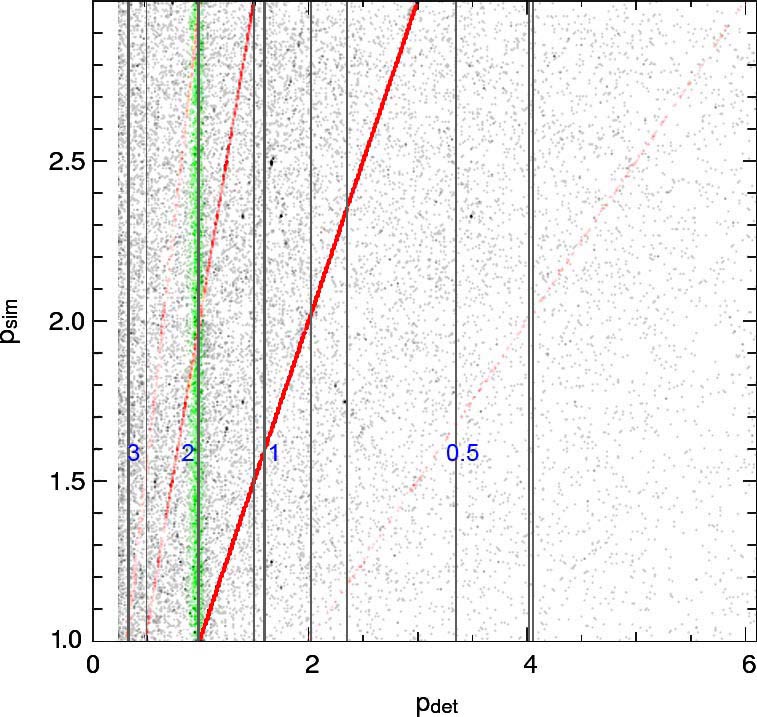}
  \caption[p-noalias]{Density plot of simulated period $p_{sim}$ against detected period $p_{det}$ for Jovian planets with periods between 1-3 days after application of our alias removal. Marked in red are the lines for correct period identification and corresponding aliases (blue number) or half, double and triple the simulated period. Further marked in green is a period area with a high amount of false detection contaminations, removed period regions (see Tab. \ref{tab:alias}) are marked as horizontal grey lines.}
  \label{fig:p-noalias}
  \end{figure}

As the next step we attempt to recover the simulated signals with our transit detection pipeline. 
As for the survey, we select the 4 best periods with highest S/N for every source. 
We remove results close to alias periods introduced by
the window function of the observing strategy (see table \ref{tab:alias}). 
\begin{table}[hbt]
\begin{tabular}{c}
 \hline\hline
  Excluded alias periods  \\
 \hline 0.315-0.335 days   \\ 
  0.498-0.500 days   \\
  0.991-1.004 days    \\ 
  1.586-1.594 days   \\ 
  1.594-1.600 days    \\ 
  1.965-1.975 days   \\ 
  2.039-2.045 days   \\ 
  2.359-2.360 days   \\ 
  3.370-3.378 days   \\ 
  4.022-4.030 days   \\ 
  4.078-4.088 days   \\ 
 \hline
 \end{tabular} 
 \caption{List of excluded alias periods that are common for false detections. We identified those periods as peaks in the abs(p$_{sim}$-p$_{det}$)/p$_{sim}$ histogram.}
 \label{tab:alias}
 \end{table} 
 Most alias cuts are not directly around harmonics of 1 day but instead slightly lower periods due the observation characteristic of seasonal change and large time gaps. Fig. \ref{fig:alias-period} shows the cut that we used for the alias period of 1 day. 
 Out of the remaining folded light curves we keep the one with the best $\chi^2$ fit.  
 In order to examine whether we could successfully recover the signal, we compare the detected period $p_{det}$ to the simulated period $p_{sim}$. This is the most reliable way of judging whether the detection was successful or not and has been utilized by other surveys as well (\citet{2013MNRAS.433..889K}, \citet{2013A&A...560A..92Z}). The low number of data points in some light curves makes the false detection of an harmonic of the period not unlikely. We accept a period deviation of 0.02\%, as shown in Fig. \ref{fig:hist-p}, and harmonics of $p_{sim}$ with orders of 0.5, 2 and 3 and following period deviations: 
 \begin{equation}\frac{p_{sim}}{p_{det}}=0.5 \pm 0.0001.\end{equation}
 \begin{equation}\frac{p_{sim}}{p_{det}}=2 \pm 0.0001,\end{equation}
 \begin{equation}\frac{p_{sim}}{p_{det}}=3 \pm 0.00015,\end{equation}
 \\A density plot of simulated against detected period is shown in Fig. \ref{fig:p-noalias}. One can see the secondary period peaks as diagonal streaks. However, any other harmonic periods are overshadowed by random detections.
 We disregard those other harmonics (e.g. 0.33 or 4) in order to keep the contamination by false-positive identifications low.
  \\
Marked in the same Fig. is a region around 1 day that shows an increased number of detections but is outside of our clipping limits, marked in green.
While we remove the large peak around 1.00 days, shown in Fig. \ref{fig:alias-period}, we cannot completely remove the area between about 0.9 to 1.1 days since that would result in too many actual transits being clipped out.
\\With a sample of more than 4 million light curves overall and more than  60000 M dwarfs, it is necessary to eliminate a large amount of light curves before visual inspection. Many surveys use a S/N criterion for preselection, however, this can be improved upon. 
\\
We take a set of simulated light curves, correct periods already selected, and set up the unmodified set of light curves as the training sample. Before starting the simulation, we remove our planetary candidates from the list of simulation targets. We take the reasonable assumption that even if there is a remaining undiscovered planetary signal in the sample, the effect will be negligible since the set consists of more than 60000 light curves. 
\\We optimize the selection criteria that we then use on the real data. Using the same approach as \citet{2013A&A...560A..92Z}, we set up a grid of over 100000 possible combinations of parameters, including S/N, transit depth, transit v shape and transit duration. We settle on the criteria that are shown in table \ref{tab:criteria} in Sect. \ref{status}. 
Besides S/N, criteria for the number of points in the transit, to rule out random noise detections, and criteria for transit duration and depth, to filter out obvious eclipsing binaries, have shown to be very effective in reducing the number of false detections.
\\As a last step we account for the visual selection bias. A signal that has been detected with the correct period and passed all of the selection criteria could still be disregarded in our visual inspection in case of only a partially visible transit. 
We implement a visual bias filter that eliminates folded transit light curves that show gaps during the eclipse, something which would lead us to dismiss the candidate in the real sample. In order to optimize this algorithm, we preselect an arbitrary set of 200 planet-injected light curves with periods of 1-10\,days. We mark those that we would accept and those we would rule out and then recreate those results with our automatic filter. This visual bias filter removes about 8\% of the remaining light curves.
\\We optimize this process for a number of 60 remaining light curves per field while recovering as many simulated objects as possible. The results are shown in table \ref{tab:criteria}. This number is the best compromise, based on our simulations. Decreasing it will impact our detection efficiency while increasing it will not give us additional detections.
\section{SURVEY STATUS} 
\label{status}
The complete data reduction for Pan-Planets has been finished and all light curves have been created as described in detail in Sect. \ref{datareduction}. We use our trapezoidal box fitting algorithm to identify the best planet candidates in our data. As described in the previous Sect., we use the simulated data for optimizing the selection criteria in order to retain about 60 light curves per field for visual inspection. The impact of each criterion for the M dwarf sample is listed in table \ref{tab:criteria}.
\begin{table}[hbt]
\begin{tabular}{c c c c}
\hline\hline
Criterion & Remaining & Removed & Change   \\
\hline  
Input & 65258& - & -  \\ 
Alias clipping & 57054 & 8204 & -12.6\%  \\ 
  S/N $\geq$ 12 & 5490 & 51564 &-90.4\%   \\
  Transit points $\geq$ 15& 5072 & 418 &-7.61\%   \\ 
  Transit duration $\leq$ 0.1& 599 & 4473 &-88.2\%   \\ 
  Transit depth $\leq$ 0.15&  553 & 46 &-7.76\%    \\
  Transit v shape $\leq$ 0.7 & 535& 18 &-3.10\%    \\
  Secondary S/N $\leq$ 10 & 419 & 116 &-21.8\% \\
\hline
\end{tabular} 
\caption{List of selection criteria and their impact on the M dwarf light curve signal detection. S/N and transit duration served as criteria with the highest impact, the former for eliminating false detections from random noise patterns, the latter for separation from binaries.}
\label{tab:criteria}
\end{table}
\\Examples for several categories of interest are listed below in Sect. \ref{cand}. Overall, we have 8 candidates around M stars, mostly hot Jupiters, 10 additional candidates around hotter stellar types, more than 300 M dwarf binary systems and 11 white dwarf variable systems. 
\subsection{Follow-up}
We are in the process of following up our candidates through a variety of observatories. Since this process is ongoing, we show only an exemplary candidate per category in the next Sect. For low-resolution follow-up, we obtained spectroscopic data from the Hobby Eberly Telescope 9.2m Low Resolution Spectrograph (HET LRS) \citep{1998SPIE.3352...34R,1998AAS...193.1003H}, Calar Alto 2.2m CAFOS \citep{2011A&A...529A..57P} and McDonald 2.1m ES2 spectrograph \citep{2014acm..conf..446R}. The data are being processed and a sample spectrum has been shown in Sect. \ref{SED-example}. We use the low-resolution spectra to characterize the host stars and compare the results to the SED fitting predictions. Further, we use the data to rule out binary stars, which possess a radial velocity amplitude that is measurable even in low resolution spectra.
\\All planet candidates are being observed during their predicted transit phase with the new 2m Fraunhofer Telescope Wendelstein \citep{2014SPIE.9145E..2DH} in order to improve the period accuracy, using the wide field imager \citep{2014ExA....38..213K}. The high-precision photometric data also allows us to improve the fitting of the transit shape, further ruling out false detections from red noise residuals and eclipsing binaries. Our predicted transit times have excellent accuracy with a deviation of less than 15 minutes over the course of three years without additional observations. As the next stage, we will record the transits again but in a wide range of photometric bands, allowing us to gain further insight into the physical parameters of this system.
\\The final step will be high-precision radial velocity measurements to eliminate all possibilities of false-positive detections, i.e. background eclipsing binary blends or brown dwarfs, and determine the mass of the planets. We are in the process of preparing those observations for the most promising candidates.
\subsection{Candidates}
\label{cand}
\begin{figure}[!ht]
\centering
\includegraphics[width=0.96\linewidth]{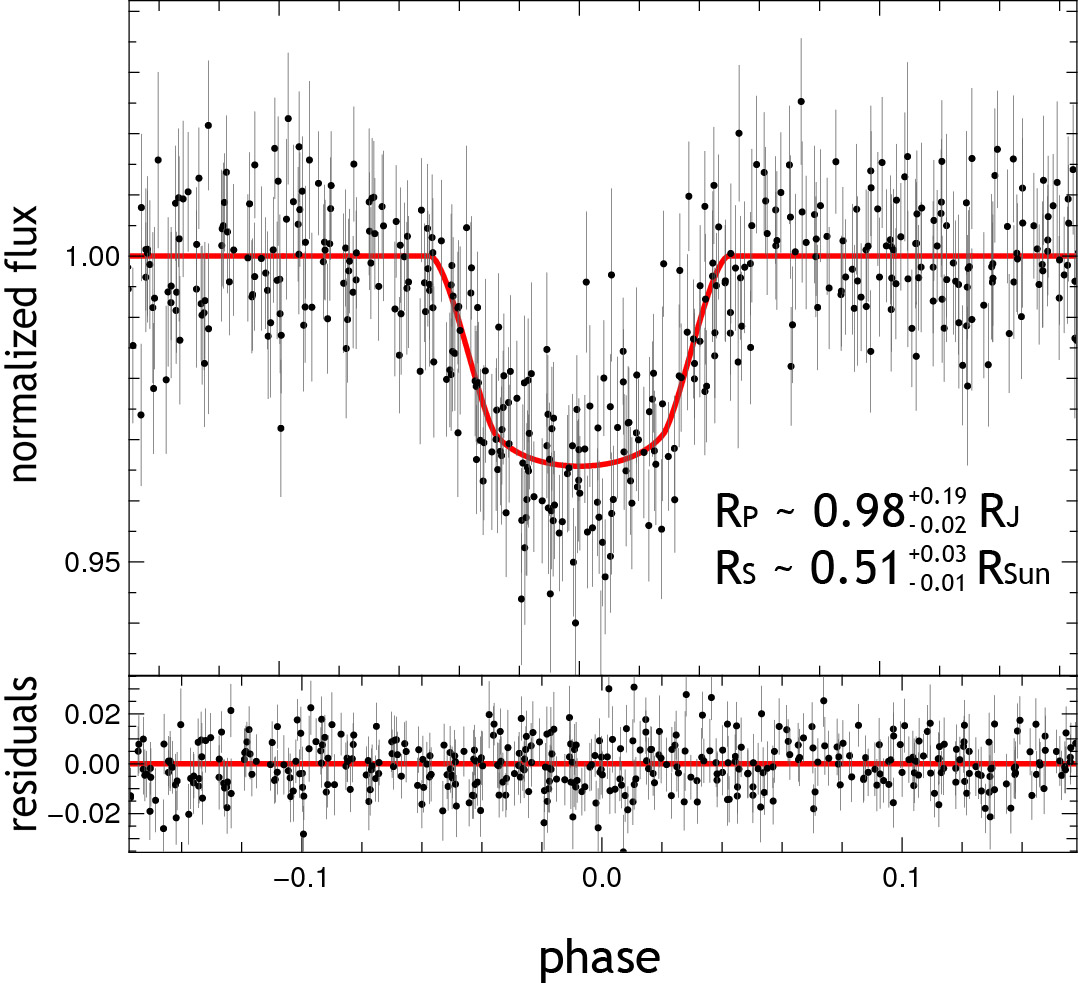}
\caption{Folded light curve (p=0.416 d) of planetary candidate 403-05317. The red line shows the best fit for parameters inclination, period, t0, planetary and stellar radii. The lower panel shows the residuals from the fit.}
\label{fig:4.03-05317}
\end{figure}

\begin{figure}[!ht]
   \centering
   \includegraphics[width=0.9\linewidth]{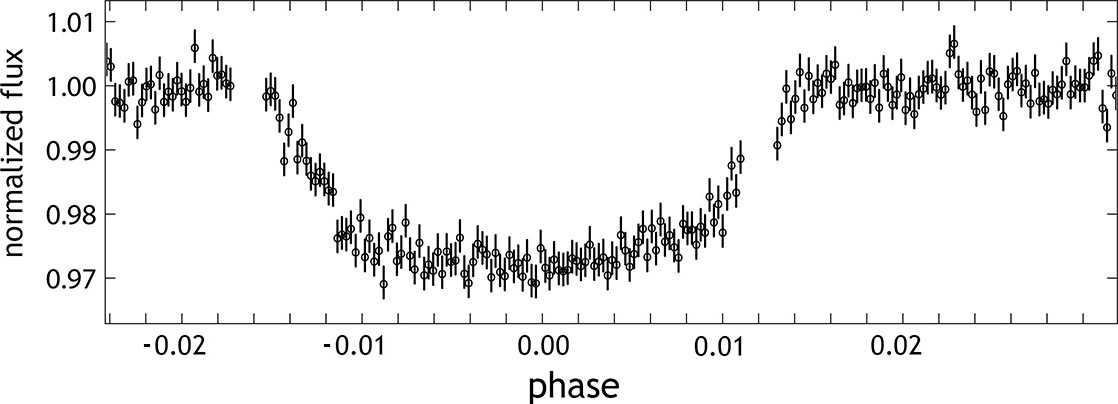}
   \caption{Transit of planetary candidate 1.40-14711, recorded with the wide field imager on the Wendelstein Fraunhofer 2m telescope in the i-band. Exposure times were 30 seconds.}
   \label{fig:1.40-14711-WST}
   \end{figure}

\begin{figure}[!ht]
   \centering
   \includegraphics[width=0.95\linewidth]{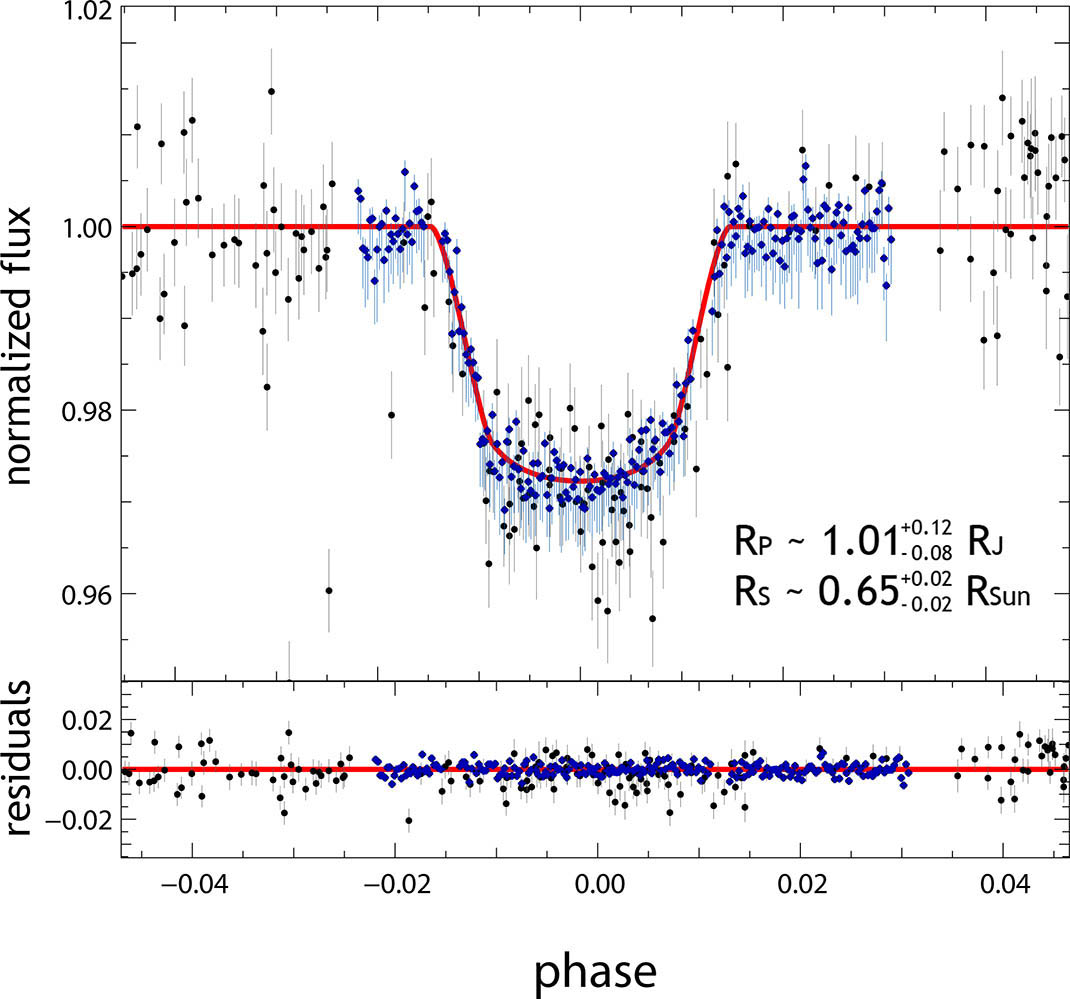}
    \caption{Folded light curve (p=2.663 d) of planetary candidate 1.40-14711. The red line shows the best fit for parameters inclination, period, t0, planetary and stellar radii. The lower panel shows the residuals from the fit. The fit includes the additional data that were taken with the 2m Fraunhofer Telescope Wendelstein (blue diamonds) besides the original Pan-Planets data (black circles).}
   \label{fig:1.40-14711}
   \end{figure}

After the identification of the planet candidates with the trapezoidal box fitting, we perform a more comprehensive fit. We determine limb darkening parameters \citep{2011A&A...529A..75C} from SED fitting and subsequently fit planetary/stellar radius, period, t0 and inclination  with a Monte-Carlo approach, further taking additional observations from Wendelstein 2m \citep{2014SPIE.9145E..2DH} into account. We display an exemplary candidate for each of the four primary categories of interest: hot Jupiters around M dwarfs, hot Jupiters around main-sequence stars, M-dwarf binary systems and other variable systems of interest.
\\Candidate 4.03-05317, shown in Fig. \ref{fig:4.03-05317}, is one of the prime planet candidates for follow-up analysis. The host star seems to be a M0 dwarf with an effective temperature of about 3950\,K and a radius of about 0.55\,$R_\odot$, which is in good agreement with the best fit for the transit shape. With radius estimates between 0.96 and 1.17\,R$_J$ and an extremely short period of 0.416\,d, it is quite uncommon and has a closer orbit than all known hot Jupiters. 
\\Candidate 1.40-14711, shown in Fig. \ref{fig:1.40-14711} and \ref{fig:1.40-14711-WST}, exhibits a clean transit signal with the most likely scenario being a hot Jupiter that is transiting in front of a late K type star with an effective temperature of about 4200\,K. We followed up this candidate with the 2m Fraunhofer Telescope Wendelstein \citep{2014SPIE.9145E..2DH}, using the wide field imager \citep{2014ExA....38..213K}. One of the resulting light curves is shown in Fig. \ref{fig:1.40-14711-WST}.
 \begin{figure}[!ht]
 \centering
 \includegraphics[width=0.9\linewidth]{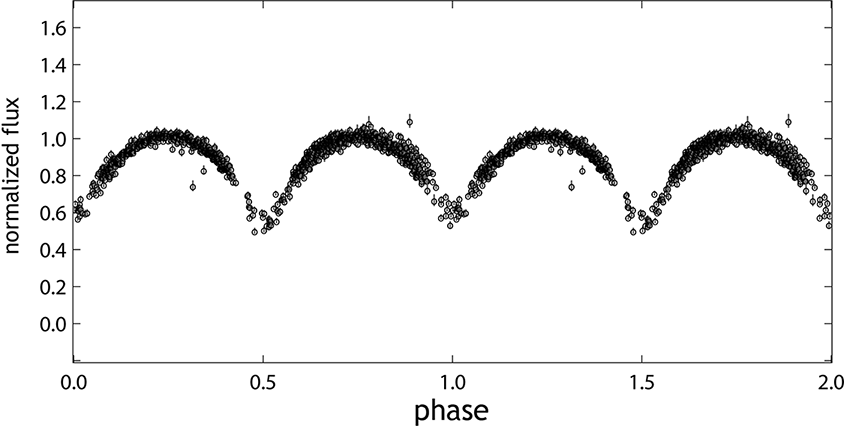}
  \caption{Folded light curve (p=0.23 d) of eclipsing binary 134-18802. The period of this binary system is extremely short (compare \citet{2011A&A...528A..90N}). The system likely consists of two similar-sized M dwarfs that are semi-detached. Note that the light curve is displayed over two phases for better visibility of the features.}
 \label{fig:1.34-18802}
 \end{figure}
 \\Candidate 1.34-18802, shown in Fig. \ref{fig:1.34-18802}, is one of many short-period (p=0.23 d) eclipsing M dwarf binary systems that we found. With light drops of 43\% and 41\% for the primary and secondary eclipse, respectively, the two members of this system seem to be of similar size and about M1 spectral type.
 \begin{figure}[!ht]
 \centering
 \includegraphics[width=0.9\linewidth]{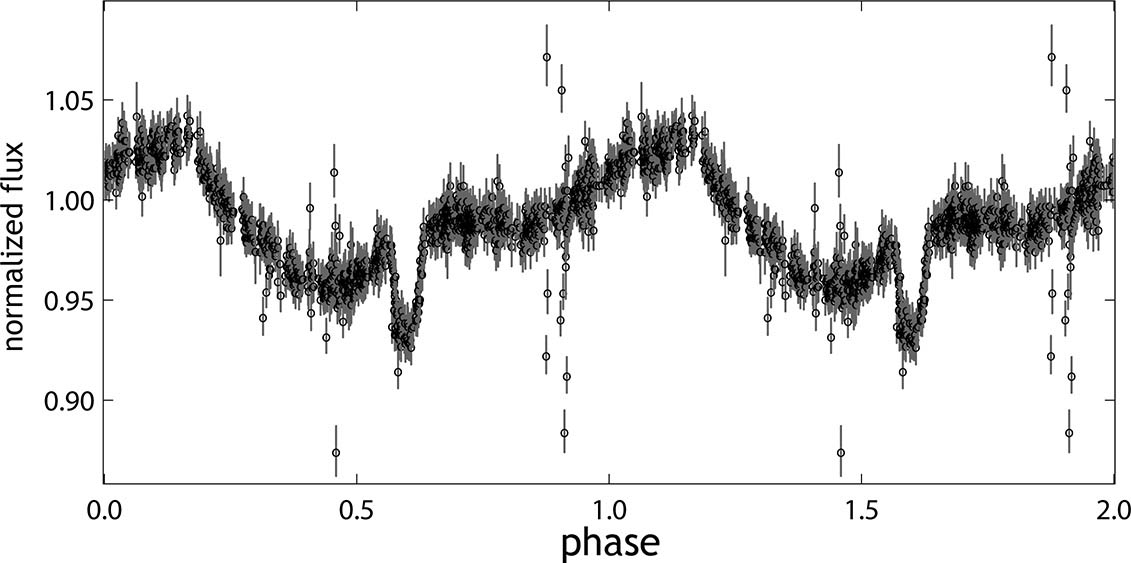}
  \caption{Folded light curve (p=2.633 d) of the unusual variable system 0.50-06948. The eclipse does not occur at phase 0.5 of the somewhat sinusoidal variation. Note that the light curve is displayed over two phases for better visibility of the features.}
 \label{fig:050-06948}
 \end{figure}
 \\Candidate 0.50-06948, shown in Fig. \ref{fig:050-06948}, is a remarkable variable system that was found in the Pan-Planets data. There are strong periodic variations, coherent over the course of four years, with an eclipse event located offset from the minimum of the variation. SED fitting predicts an high effective temperature of about 10000\,K. We recorded a spectrum with ES2 spectrograph at the McDonald observatory, confirming that the system exhibits broad Balmer-lines with an otherwise continuous spectrum. As of now the exact nature of this system is unclear. 
\section{Discussion}
\label{discussion}
\subsection{Detection Efficiency}
\begin{figure}[!ht]
 \centering
 \includegraphics[width=0.8\linewidth]{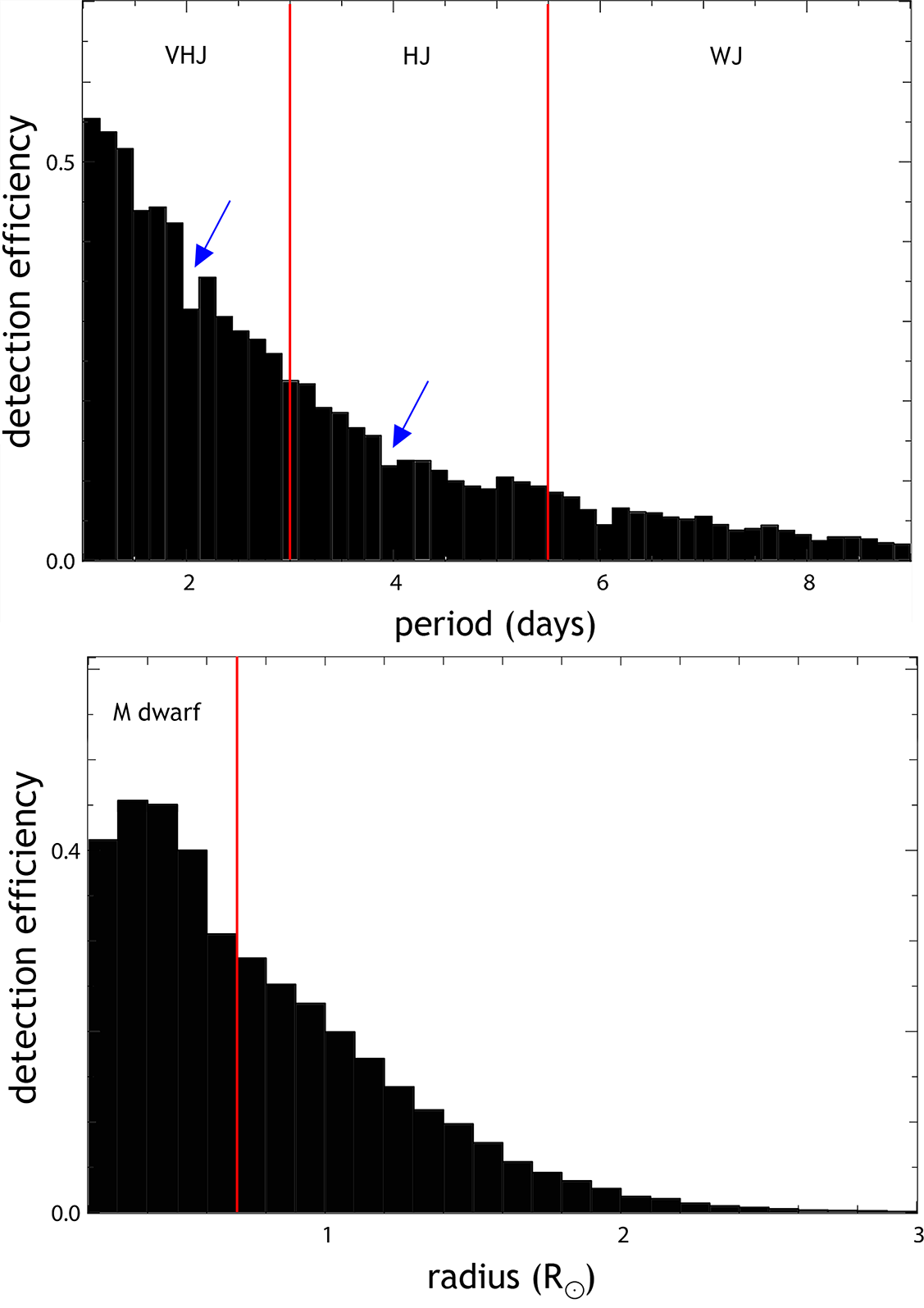}
 \caption[efficiency]{Top:
 Detected period against detection efficiency for all hot Jupiter populations around M dwarfs (divided by red lines). One can see two gaps at 1.6 and 4.0 days, resulting from our alias detection removal (blue arrows).
 Bottom: Histogram of stellar radius against detection efficiency. We combined the results from the M dwarf VHJ simulation and the VHJ simulations for hotter dwarf stars (divided by red line).}
 \label{fig:efficiency}
 \end{figure}

\begin{figure}[h]
 \centering
 \includegraphics[width=\linewidth]{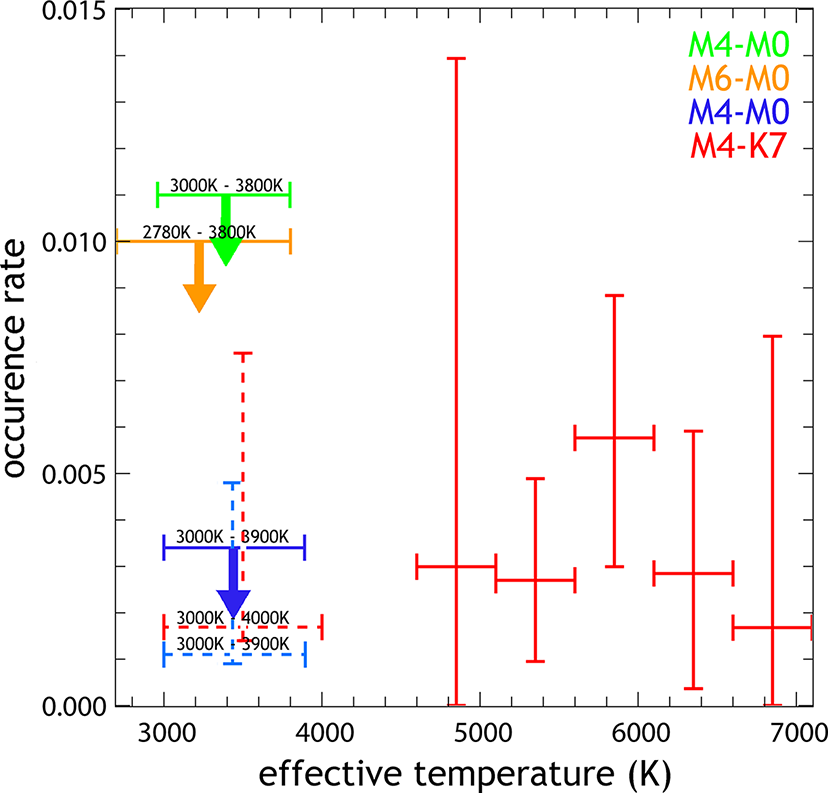}
 \caption{Adaptation of Fig. 13 in \citet{2013MNRAS.433..889K}, showing the hot Jupiter fractions determined by different surveys. We added our new results, marked in dark blue (upper limit) and light blue dotted line (fraction in case of a successful detection). Orange shows the limits derived from radial velocity surveys \citep{2013A&A...549A.109B}, red from the Kepler survey (extracted by \citet{2013MNRAS.433..889K}), red in dotted lines from our own simulations for Kepler and green from the WFCAM transit survey \citep{2013A&A...560A..92Z}.}
 \label{fig:fraction}
 \end{figure}

For each planet population, we repeat 100 simulation runs per M dwarf and 40 runs per FGK star. This adds up to 50 million individual runs per planet population for M dwarfs and 245 million runs for the K, G and F star population. We end up with a recovery ratio for the individual planetary populations shown in table \ref{tab:results}. One can see that the detection efficiency is increasing strongly for lower periods and larger radii. A histogram of the detection efficiency against the period for M dwarfs can be seen in Fig. \ref{fig:efficiency} on the top panel. 
\begin{table}[hbt]
\begin{tabular}{c|c c c}
\hline\hline
M dwarf & Simulated & Recovered & Efficiency \\
\hline  VHJ  & 772870 & 45.6\%& 40.6\%  \\ 
  HJ   & 471484 & 17.5\% &   14.5\% \\
  WJ  & 198983 & 7.3\% &  5.45\%   \\ 
  VHS  & 772044 & 19.8\% &  18.5\%   \\ 
  VHN  & 767916 &  10.3\% &  9.68\% \\ 
\hline
K, G, F  dwarf\\
\hline  VHJ & 2476929 & 9.32\% & 8.72\%  \\ 
\hline
\end{tabular} 
\caption{Detection efficiencies for different planet populations. For M dwarfs, we use all 65258 targets minus the 8 planet candidates, for KGF dwarfs we use 460910 targets, excluding the planet candidates. The percentage of recovered planets is calculated by normalizing the number of correct detections by the number of possible detections (see Eq. \ref{eq:sim}). The efficiency is determined after selection criteria, alias clipping and visual bias filter have been applied to the results from the BLS analysis. For smaller stellar radii, larger planetary radii and shorter periods, the efficiency is higher.}
\label{tab:results}
\end{table} 
\\One has to keep in mind that the recovery efficiencies shown in table \ref{tab:results} include possible cases of barely observable transits - even the slightest overlaps between planet and star are being simulated where the transit would take place within only a few seconds as an extreme example. There further are simulated light curves that are not observable due to data gaps or badly timed transits that constantly fall outside of our observing windows. Even with perfect data, it would therefore be hard to reach 100\% detection efficiency. The detection efficiency in relation to the stellar radius is shown in the bottom panel of Fig. \ref{fig:efficiency}. It is clear that the stellar radius has a significant impact on the detection rate. The efficiency strongly decreases after 0.5 R$_\odot$. Since the efficiency reaches a plateau before that, we assume that this is the maximum achievable detection efficiency with Pan-Planets. The other transit signals may be lost in observation gaps or in strong stellar variability that masks the signal and cannot be properly distinguished due to an insufficient number of data points.
For K, F and G dwarfs combined, we expect to detect $3.0^{+3.3}_{-1.6}$ transiting VHJs, assuming an occurrence rate of $0.1408\cdot(1^{+1.1}_{-0.54})\%$  based on the OGLE-III transit search \citep{2006AcA....56....1G}. 
\\Our large sample means that, assuming a null result in which none of the M dwarf candidates turn out to be actual planets, we can set new upper limits for the planetary occurrence rates of hot Jupiters around those stars. The number of detections is characterized with a Poisson distribution. Therefore, assuming a number of k planets in our sample, the probability of having N$_{det}$ planets is 
\begin{equation}
P_k = \frac{{N_{det}^k}}{k!}  e^{-N_{det}}.
\end{equation}
The geometric probability for a visible transit is empirically being accounted for in our simulations. For hot Jupiters, between 9.8\% ($\mathrm{1\,d \leq p \leq 3\,d}$) to 2.5\%   ($\mathrm{5\,d \leq p \leq 10\,d}$) of the simulated transits pass our visibility criterion (Eq. \ref{eq:sim}). The detection efficiency is therefore a combination of the geometric probability $P_{transit}$ and the detector efficiency $P_{det}$. We now assume the null result, e.g. $k=0$. In order to compare our results to \citet{2013MNRAS.433..889K} and \citet{2013A&A...560A..92Z}, we also use a confidence interval of 95\%. Solving
\begin{equation}
\label{eq:poisson}
P(N_{det} < N_{max}) = \int\limits_{0}^{N_{max}} e^{-N}dN = 0.95,
\end{equation}
we get N$_{max}$=3. We can calculate the upper limit by replacing the number of observed planets N$_{det}$ with the product of the number of stars with the detection efficiency and fraction $f$ so that $N_{det}=N_{stars}\cdot P_{det}\cdot P_{transit}\cdot f$:
\begin{equation}
\label{eq:frac}
f_{95\%} \leq \frac{3}{N_{stars}\cdot P_{det}\cdot P_{transit}}.
\end{equation}
Taking the individual detection efficiencies in every field, the geometric probability for each period bin and assuming that the distribution of planetary radii is even into account, we end up with an upper limit of 0.34\%. This is a significantly lower result than found in previous surveys where small sample sizes counteracted higher detection efficiencies. Splitting up the results for M0-M2 and M2-M4 sub groups as done in \citet{2013MNRAS.433..889K} and \citet{2013A&A...560A..92Z}, we derive upper limits of 0.49\% and 1.1\%, respectively.
However, we possess several plausible M dwarf hot Jupiter candidates. 
Assuming one correctly identified hot Jupiter, we calculate the occurrence rate to be 0.11$^{+0.37}_{-0.02} $\% with a 95\% confidence limit. For the upper uncertainty, we integrate equation \ref{eq:poisson} in the range of 1 to $N_{max}$ and determine the fraction limit. For the lower uncertainty, we consider the scatter of our simulations and calculate the difference between the average and minimum detected planets per simulation run. It may look counter-intuitive that a successful detection lowers the supposed fraction. One has to keep in mind that the null result describes the upper limit while a successful detection allows for an estimate of the fraction. Additionally, the uncertainties of the fraction estimate are higher than the null result's limit.  
As another comparison, we determine the best-case results from the Kepler survey. We assume a number of 3897 stars in the temperature range between 3000\,K and 4000\,K in the distribution of \citet{2013ApJ...767...95D} and simulate a system with the same hot Jupiter population as ours for the given stellar radius. If the criterion 
\begin{equation}
sin(i)< \frac{R_{star}-R_{planet}}{a}, 
\end{equation}
e.g. a full transit of the planet, is met, we assume a successful detection due to the photometric accuracy and the long baseline of Kepler. Note that this criterion is different to eq. \ref{eq:sim}: here we assume a full eclipse of the planet. The resulting fraction with one confirmed planet \citep{2012AJ....143..111J} is $0.17(^{+0.67}_{-0.04})$\%, an occurrence rate that is on par with our own results although 50\% higher and with a larger error due to the small sample of cool Kepler stars. Further, the inclusion of stars up to 4000\,K means that this fraction cannot be compared directly. Furthermore, there are three additional hot Jupiter candidates in the Kepler database\footnote{http://exoplanetarchive.ipac.caltech.edu/}, KOIs 3749.01, 1654.01 and 1176.01. All of their radii are very close to that of Jupiter and show no signs of inflation, e.g. radii much larger than 1\,$\mathrm{R_J}$ that is frequent for hot Jupiters. It is possible that they are in fact Brown Dwarfs, so further follow-up will be necessary to determine their true nature. This means that Kepler's occurrence rate limits might end up being higher than assumed here, depending on whether or not all of the remaining Kepler candidates are planets.
\\We illustrate the impact of this new occurrence limit in Fig. \ref{fig:fraction}. Our result pushes the upper limit down to the level of other main-sequence stars. However, theoretical models \citep{2005PThPS.158...68I,2010PASP..122..905J,2012A&A...541A..97M} point to an even lower fraction. 
\subsection{Comparison to the expected number of detections}
When comparing our measured detection efficiency to the predictions of
\citet{2009A&A...494..707K}, one has first to consider the difference
in the number of data points per star.  Pan-Planets was scheduled for
4\% of the total observing time, which we actually received. However,
\citet{2009A&A...494..707K} assumed that this would add up to 280 h,
while in the end we received 165 h because of different reasons
(delayed fully operational readiness, weather, maintenance).  This
significantly decreased the detection efficiency. The change in
observing time was shown to have a non-linear impact, e.g. doubling
the amount of observing time increased the number of detected planets
by a factor of three (see tables 8 and 9 in
\citet{2009A&A...494..707K}) for periods longer than 3
days. Furthermore, while we assumed a precision up to 4 mmag red noise
residual, the majority of light curves now has a precision between
5-10 mmag. There is no directly comparable simulation, so we would
have to adjust the previous red noise models. The unforeseen issues
for bright stars in 2010 could also not have been taken into account,
meaning that there are less than 1500 data points for any bright
source (i' $\leq$ 15.5 mag). \\We scale down table 8 in
\citet{2009A&A...494..707K}, which assumes 120 hours of data taken in
one year, for the aforementioned effects - a red noise residual level
of 5\,mmag and fewer data points than previously assumed. We find our
expected number of $3.0^{+3.3}_{-1.6}$ detected hot Jupiters to be consistent with the scaled estimate of 7.4$\pm$2.9
detections.
\section{CONCLUSION}
\label{conclusions}
In the years 2010-2012, the Pan-Planets survey observed seven
overlapping fields in the Galactic disk for about 165 hours. The main
scientific goal of the project is to find transiting planets around M
dwarfs, however, with more than 4 million sources brighter than
i'=18 in the 42 sq. deg. survey area the data are a valuable
source for a diversity of scientific research.\\
We established an efficient procedure to determine the stellar
parameters $\mathrm{T_{eff}}$ and log\,$g$ of all sources using a method based on
SED fitting, utilizing a three-dimensional dust map and proper motion information. In this way we were able to
identify more than 65\,000 M dwarfs which is by far the biggest sample
of low-mass stars observed in a transit survey up to now.\\
Using a optimized difference imaging data processing pipeline we
reached a photometric precision of 5\,mmag at the bright end at around
$i_{P1}$ = 15\,mag. This makes Pan-Planets sensitive to short period
hot Jupiters and hot Neptunes around M dwarfs and short period hot
Jupiters around hotter stellar types.\\
To search for planetary transits we used a modified BLS algorithm. We
applied several selection criteria which have been optimized using
Monte-Carlo simulations in order to reduce the number of visually
inspected light curve from several million down to a about 60 per field. We detected several planet candidates around M dwarfs and
hotter stars which are currently being followed up. In addition, we
found many interesting low-mass eclipsing binaries and eclipsing white
dwarf systems which we will study in detail in the current observing
season.\\
Using Monte-Carlo simulations we determined the detection
efficiency of the Pan-Planets survey for several stellar and planetary
populations. We expect to find  $3.0^{+3.3}_{-1.6}$ hot Jupiters around F, G
and K dwarfs with periods lower than 10 days based on the planet
occurrence rates derived in previous surveys. For M dwarfs, the fraction
of stars with a hot Jupiter is under debate. With the large sample
size of Pan-Planets, we were able to determine a planet fraction of
 $0.11(^{+0.37}_{-0.02})$\% in case one of our candidates turns out to 
 be a real detection. For this result, we considered the average detection rate of the simulations and compared the scatter at a 95\% confidence.\\
If however none of our candidates is real, we were able to put a 95\%
confidence upper limit of 0.34\% on the hot Jupiter occurrence
rate of M dwarfs. This limit is higher than the calculated fraction in case of a successful detection, however, the uncertainties of the fraction are in turn higher than this upper limit.  This result is a significant improvement over
previous estimates where the lowest limit published so far is 1.1\%,
found in the WTS survey \citep{2013A&A...560A..92Z}, or, using our approach to estimate the generous best case for Kepler, $0.17(^{+0.67}_{-0.04})$\%. Despite the
significant improvement, our upper limit is still comparable to the
occurrence rate of hot Jupiters around F, G and K dwarfs, even more so in case of a successful detection. The estimates from \citet{2006AcA....56....1G}  based on the OGLE-III transit search seem to be in good agreement with our new limits. Therefore we
could not yet confirm the theoretical prediction of a lower rate for cool
stars. Other surveys with even larger M dwarf samples and/or better
detection efficiency will be needed to answer this question.\\
\begin{acknowledgements}
\\The Pan-STARRS1 Surveys (PS1) have been made possible through contributions of the Institute for Astronomy, the University of Hawaii, the Pan-STARRS Project Office, the Max-Planck Society and its participating institutes, the Max Planck Institute for Astronomy, Heidelberg and the Max Planck Institute for Extraterrestrial Physics, Garching, The Johns Hopkins University, Durham University, the University of Edinburgh, Queen's University Belfast, the Harvard-Smithsonian Center for Astrophysics, the Las Cumbres Observatory Global Telescope Network Incorporated, the National Central University of Taiwan, the Space Telescope Science Institute, the National Aeronautics and Space Administration under Grant No. NNX08AR22G issued through the Planetary Science Division of the NASA Science Mission Directorate, the National Science Foundation under Grant No. AST-1238877, the University of Maryland, and Eotvos Lorand University (ELTE).
\\This paper contains data obtained with the 2m Fraunhofer
Telescope of the Wendelstein observatory of the Ludwig-Maximilians 
University Munich.
\\We thank the staff of the Wendelstein observatory for technical help and strong support, including observing targets for us, during the data acquisition.
\\This publication makes use of data products from the Two Micron All Sky Survey, which is a joint project of the University of Massachusetts and the Infrared Processing and Analysis Center/California Institute of Technology, funded by the National Aeronautics and Space Administration and the National Science Foundation.
\\This paper includes data taken at The McDonald Observatory of The University of Texas at Austin.
\\The Hobby-Eberly Telescope (HET) is a joint project of the University of Texas at Austin, the Pennsylvania State University, Stanford University, Ludwig-Maximilians-Universität München, and Georg-August-Universität Göttingen. The HET is named in honor of its principal benefactors, William P. Hobby and Robert E. Eberly.
\\The Marcario Low Resolution Spectrograph is named for Mike Marcario of High Lonesome Optics who fabricated several optics for the instrument but died before its completion. The LRS is a joint project of the Hobby-Eberly Telescope partnership and the Instituto de Astronomía de la Universidad Nacional Autónoma de México.
\end{acknowledgements}
\bibpunct{(}{)}{;}{a}{}{,} 
\bibliographystyle{aa} 
\bibliography{bibliography} 

\end{document}